\documentclass[usenatbib,twoside]{mn3e} 
\usepackage{amsmath,amssymb}
\usepackage[figuresright]{rotating}
\usepackage{tablefootnote}   
\usepackage{graphicx}         
\usepackage{gensymb}        
\usepackage{color}               
\usepackage{natbib}
\usepackage{times}
\usepackage{fixltx2e} 
\usepackage[flushleft]{threeparttable}
\usepackage{caption}
\usepackage{subfig}

\newcommand{\ramses}{{\sc ramses}}

\newcommand{\Msol}{\,{\rm M}_\odot} 
 
\newcommand{\Gpc} {{\,\rm Gpc}}
\newcommand{\Mpc} {{\,\rm Mpc}}
\newcommand{\kpc} {{\,\rm kpc}}
\newcommand{\pc} {{\,\rm pc}} 
 
\newcommand{\cc}{{\,\rm {cm^{-3}}}}

\newcommand{\kmsec}{{\,\rm {km\,s^{-1}} }}

\def\Myr{\,{\rm Myr}}
\def\yr{\,{\rm yr}}
\def\um{\,{\rm \mu m}}

\newcommand{\diff}{{\rm d}}
\newcommand{\ang}{{\,\rm\AA}}
\newcommand{\ev}{{\,\rm eV}}
\newcommand{\lunit}{{\,\rm erg\,s^{-1}\,cm^{-2}}}

\newcommand{\lunittl}{{\,\rm erg\,s^{-1}\,cm^{-2}\,\ang^{-1}\,arcsec^{-2}}}
\newcommand{\mas}{{\,\rm mas}}
\newcommand{\flya}{{f_{\rm Ly\alpha}}}

\newcommand{\piii}{Pop. III }
\newcommand{\Piii}{Pop. III}
\newcommand{\pii}{Pop. II }

\newcommand{\pl}{Pop. I }

\newcommand{\he}{{\rm He{\small II}} }

\newcommand{\heii}{{\rm He{\small II}}$\lambda1640$ }
\newcommand{\Heii}{{\rm He{\small II}}$\lambda1640$}
\newcommand{\ha}{\rm H$\alpha$ }
\newcommand{\Ha}{\rm H$\alpha$} 
\newcommand{\lya}{\rm Ly$\alpha$ }
\newcommand{\Lya}{\rm Ly$\alpha$} 
\newcommand{\eelt}{ELT }
\newcommand{\Eelt}{ELT}
\newcommand{\sfr}{\rm SFR}
\newcommand{\cloudy}{{\sc cloudy} }
\newcommand{\cloudyp}{{\sc cloudy}}
\newcommand{\newhorizon}{{\sc NewHorizon} }
\newcommand{\newhorizonp}{{\sc NewHorizon}}
\newcommand{\hsim}{{\sc hsim} }
\newcommand{\hsimp}{{\sc hsim}}

\title[\piii Stars and HARMONI] {Predicting the Observability of Population III Stars with ELT-HARMONI via the Helium $1640\ang$ emission line}\author[Kearn Grisdale
  et al.] {\parbox[t]{\textwidth}{Kearn Grisdale$^1$\thanks{kearn.grisdale@physics.ox.ac.uk}, Niranjan Thatte$^{1}$, Julien Devriendt$^{1}$, Miguel Pereira-Santaella$^{1,2}$,  Adrianne Slyz$^{1}$, Taysun Kimm$^{3}$ , Yohan Dubois$^{4}$ and Sukyoung K. Yi$^{3}$}\vspace*{6pt}\\
  	$^{1}$ Sub-department of Astrophysics, University of Oxford, Keble Road, Oxford OX1 3RH\\
	$^{2}$ Centro de Astrobiolog\'{a} (CSIC-INTA), Ctra. de Ajalvir, Km 4, 28850, Torrej\'{o}n de Ardoz, Madrid, Spain\\
	$^{3}$ Department of Astronomy, Yonsei University, 50 Yonsei-ro, Seodaemun-gu, Seoul 03722, Republic of Korea\\
	$^{4}$ Institut d'Astrophysique de Paris, UMR 7095, CNRS, UPMC Univ. Paris VI, 98 bis boulevard Arago, 75014 Paris, France
	}
\date{\today}
\begin{document}
\maketitle
\graphicspath{ {Figures/} }
\begin{abstract} 
	Population III (\Piii) stars, as of yet, have not been detected, however as we move into the era of extremely large telescopes this is likely to change. One likely tracer for \piii stars is the \heii emission line, which will be detectable by the HARMONI spectrograph on the European Extremely Large Telescope (\Eelt) over a broad range of redshifts ($2\leq z\leq14$). By  post-processing  galaxies from the cosmological, AMR-hydrodynamical simulation \newhorizon with theoretical spectral energy distributions (SED) for Pop. III stars and radiative transfer (i.e. the Yggdrasil Models and \cloudy look-up tables respectively) we are able to compute the flux of \heii for individual galaxies. From mock 10 hour observations of these galaxies we show that HARMONI will be able to detect \piii stars in galaxies up to $z\sim10$ provided \piii stars have a top heavy Initial Mass Function (IMF). Furthermore, we find that should \piii stars instead have an IMF similar to those of the Pop. I stars, the \heii line would only be observable for galaxies with \piii stellar masses in excess of $10^{7}\Msol$, average stellar age $<1\Myr$ at $z=4$. Finally, we are able to determine the minimal intrinsic flux required for HARMONI to detect \piii stars in a galaxy up to $z=10$.

\end{abstract}


\section{Introduction}
\label{sect:intro}

The very first stars in the Universe, known as Population III (\Piii) stars, are believed to have been extremely massive \citep[$\langle M_{\star}\rangle\geq100\Msol$, e.g. see][and references within]{Schwarzschild:1953aa,Larson:1998aa,Bromm:1999aa,Abel:2000aa}. These stars would have formed out of the primordial gas created during Big-Bang nucleosynthesis and as a result would have been chemically pristine. During the course of their lifetimes these stars would have reionized the surround gas and, perhaps more importantly, created the first batch of elements heavier than lithium in their interiors \citep{Ostriker:1996aa}. Through feedback processes (i.e. winds and radiations) and eventual supernova phase, \piii stars heat and enrich their environment which has a direct impact on the further star formation \citep{Omukai:1999aa,OShea:2005aa}.  Due to this enrichment it is unlikely to find \piii stars at a low redshift \citep[$z\leq3$,][]{Johnson:2010aa}. Furthermore, as a result of their (predicted) mass these stars would have relatively short lifetimes, on the order of $1$-$10\Myr$ \citep{Schaerer:2002aa} and thus are only expected to be found at high redshifts ($z\gg2$). This first generation of stars therefore determined the composition of future generations (Population II and I) and changed how gas throughout the Universe was able to cool i.e. allow for metal line cooling \citep{Mackey:2003aa}. 

Due to their importance in determining the evolution of galaxies and the intergalactic medium there have been a number of attempts to observe \piii stars. 
However, detections have proven difficult in part due to the short lifetime of \piii stars (assuming $\langle M_{\star}\rangle\geq100\Msol$) which \citep[assuming they were formed at $z>2$, ][]{Dijkstra:2007aa,Nagao:2008aa,Kashikawa:2012aa}, are not expected to still be around at $z=0$. Furthermore, due to the enriched metallicity of interstellar, circumgalactic and intergalactic mediums (ISM, CGM and IGM), by definition, the local Universe is no longer able to form new \piii stars. Detection of these stars therefore requires telescopes that are able to probe sufficiently high redshifts that \piii stars are either still forming and form a large fraction of the total star formation in the galaxy at that epoch. 

When doubly ionised helium recombines and undergoes the $n=3\rightarrow2$ transition it produces a bright emission line at $1640\ang$ (\Heii).
As a result of their (theoretical) high mass, \piii stars are predicted to produce a large number of photons with sufficient energy $(\ge54.4\ev)$ to fully ionise helium in the surrounding ISM and thus \heii could be a strong indicator, if not direct tracer, for such stars \citep{Jimenez:2006aa}.
Unfortunately, other objects such as Wolf-Rayet (WR) stars and Active Galactic Nuclei (AGN) also produce the \heii emission line \citep[see][and references within]{Francis:1991aa,Leitherer:1995aa,Schaerer:1998aa,AllensAstro}. With the aid of additional diagnostics it should be possible to distinguish between these different sources. For example,  the spectra of both AGN and WR stars would normally contain emission from C III and C IV \citep[][]{Leitherer:1996aa,Reuland:2007aa,Allen:2008aa}. Additionally, \heii produced by \piii stars is expected to be relatively narrow ($<700\kmsec$) while lines from WR are expected to be significantly broader \citep[$>800\kmsec$,][]{Leitherer:1995aa,Cassata:2013aa}. 
A key diagnostic is the ratio of helium to hydrogen ionising photons, which depends on the hardness of the UV radiation field, and is directly connected to the stellar temperature, and therefore, mass.
As the \heii line is unable to ionise hydrogen, it is not heavily absorbed by the ISM \citep{Woods:2013aa} and therefore He{\small II} emissions from high redshift galaxies should be detectable.

One possible detection of a galaxy containing \piii stars was made by \cite{Sobral:2015aa} when looking for bright Lyman-$\alpha$ (\Lya) galaxies. The object they identified, designated ``CR7'', had a strong--narrow \heii emission line. Furthermore they ruled out the possibility of WR stars or AGN and concluded that CR7 is ``the strongest candidate for a \Piii-like stellar population''. With the help of additional observations, \cite{Bowler:2017aa} found that CR7 is contaminated by the {[O{\small III}]} $\lambda \lambda 4959, 5007$  doublet. They therefore argued that the detected \heii signal in CR7 is more likely to be from either a low-mass, narrow-line AGN or a young low-mass starburst (if binaries are included in the model).

Despite the current difficulties faced by observations, new facilities such as the European Extremely Large Telescope (\Eelt), with its $39 {\rm\,m}$ diameter primary mirror, 
present a new opportunities for the detection of \piii stars.  The High Angular Resolution Monolithic Optical and Near-infrared Integral field spectrograph (HARMONI), currently being built, will serve as the work-horse spectroscopic instrument on \eelt and will provide spectra from $0.47$ to $2.45\um$ \citep[][]{Thatte:2014aa}. This large wavelength range allows for the possibility that the \heii line will be detectable for $2\lesssim z\lesssim14$. Furthermore the high spatial resolution of HARMONI may also provide details about the spatial distribution of \piii stars inside their host galaxies. When combined with the large ($39{\rm\,m}$) primary mirror of the \Eelt, HARMONI is ideally suited to detect \piii stars

In this work we explore the possibility of detecting \piii stars with HARMONI. To achieve this we combine the numerical cosmological simulation: \newhorizon with the spectral synthesis code \cloudy to generate mock spectra of galaxies, containing \piii  stars for $3\lesssim z\lesssim10$, which are ``observed''  using the HARMONI simulator \hsimp. In addition we explore how the mass function of \piii stars will impact observations. The paper is organised as follows. In Section~\ref{sect:meth} we outline the process of generating mock observations. We present our results in Section~\ref{sect:results} and discuss their implications in Section~\ref{sect:disc}. Section~\ref{sect:con} summarises this work and its conclusions.

\section{Method}
\label{sect:meth}

\subsection{\newhorizon}
\label{meth:nh}
\subsubsection{Simulation Overview}

Our source for mock ``observable'' galaxies is the \newhorizon simulation, a high spatial resolution ($\Delta x\sim35\pc$\footnote{This is the linear size of the smallest cell.}), hydrodynamical, cosmological simulation run using the hydro+$N$-body, Adaptive Mesh Refinement (AMR) code {\ramses} \citep{Teyssier:2002aa}. Here we give a brief overview of the simulation but direct the reader to \cite{Dubois:2020aa} for complete details \citep[see also][]{Park:2019aa}. \newhorizon is a re-simulation of a field environment spherical region with a radius of $10\Mpc$ comoving which has been extracted from the Horizon-AGN simulation \citep{Dubois:2014aa,Kaviraj:2017aa}. 

\newhorizon started at a redshift of $z=45$ and has currently reached $z\sim0.7$ after 40 million CPU hours. 
To counter cosmological expansion, new levels of refinement are unlocked as the simulation progresses to keep the maximum resolution as close to $35\pc$ as possible. We give the actual resolution ($\Delta x_{z}$) for each snapshot that we analyse in Table~\ref{table:gals}. The simulation assumes cosmological parameters  consistent with the WMAP-7 data \citep[][]{Komatsu:2011aa}, i.e. Hubble constant $H_{0} = 70.4 \kmsec\Mpc^{-1}$, total mass density $\Omega_{m}=0.272$, total baryon density $\Omega_{b}=0.0455$, dark energy density $\Omega_{\Lambda}=0.728$, amplitude of power spectrum $\sigma_{8}=0.809$, and power spectral index $n_{s}=0.967$. Dark matter is modelled as collisionless particles, with a mass resolution of $10^{6}\Msol$. The simulation includes several important physical processes such as star formation, stellar feedback from star particles, AGN feedback from ``black hole'' sink particles, however it does not include explicit radiative transfer. Gas is able to cool to $1{\rm\,K}$ via a metal dependent collisional equilibrium model of radiative cooling \citep[][]{Sutherland:1993aa,Rosen:1995aa} in the presence of a uniform ultraviolet radiation field after the epoch of reionisation \citep[i.e. $z=10$,][]{Haardt:1996aa}. At high redshift, due to the lack of metals, cooling is primarily achieved via molecular hydrogen, however the simulation does not model molecular gas and consequently, this method of cooling has to be approximated. 
This is achieved by assuming an initial metallicity of $Z_{\rm init }=10^{-3}{\rm Z}_{\odot}$ \citep[e.g.][]{Wise:2012aa}and using metal line cooling at all times.

Star formation occurs on a cell by cell basis when their gas number density $\geq10{\rm\, cm}^{-3}$ and temperature $<2\times10^{4}{\rm\,K}$. Stars particles form according to a Schmidt law \citep{Schmidt:1959aa} with the star formation efficiency per free-fall time determined by the local thermo-turbulent conditions of the ISM \citep[][]{Kimm:2017aa}. Each star particle an initial (or birth) mass of $M_{\star}\geq10^{4}\Msol$ and is assumed to represent a population of stars which follows a Chabrier Initial Mass Function \citep[IMF, ][]{Chabrier:2005aa} with lower and upper mass cutoffs of $0.1\Msol$ and $150\Msol$ respectively. These particles in turn feed back gas \citep[$31\%$ of $M_{\star}$ with appropriate momentum, see][]{Kimm:2014aa} into the ISM  via supernovae explosions occurring $5\Myr$ after their formation. 

Cells where both gas and stellar densities are $>10{\rm\, cm}^{-3}$ can form black holes (sink particles), with an initial seed mass of $10^{4}\Msol$. These particles can then grow through Bondi--Hoyle--Lyttleton accretion \citep{Bondi:1944aa,Hoyle:1939aa} capped at the Eddington limit. Two models are employed to add feedback from AGN: radio  and quasar mode. The choice of model is set by ratio of the gas accretion rate to the Eddington limit \citep[see][to which we refer the reader for details]{Dubois:2012aa}, and in the jet mode case, the feedback efficiency depends on the spin of the black hole \citep[][]{Dubois:2014ab}. 

\subsubsection{Selection of Galaxies}
\label{meth:nh:sel}
The \newhorizon galaxy catalogue contains hundreds of galaxies at any given redshift \citep[see][]{Dubois:2020aa}.  In order to identify ideal candidates for observations we reduce the sample of galaxies by only considering those for which  $M_{\rm \star,PopIII,tot}/M_{\star,tot}\geq0.5$. Here $M_{\rm \star,PopIII,tot}$ and $M_{\star,tot}$ are the total mass in \piii stars and the total stellar mass of the galaxy respectively. Furthermore we require that the stellar half-mass radius for the \piii stars ($R_{\rm0.5,P3}$) be less than one kiloparsec for a galaxy to be considered. We determine whether or not a star particle is \piii by its metallicity. If a particle has $Z_{\rm init}< Z<Z_{\rm crit}$ we consider it to be a \piii particle, where $Z_{\rm crit}=0.02\,Z_{\odot}$\footnote{Throughout this work we adopt $Z_{\odot}=0.02$}$(=4\times10^{-4})$ adopted from \cite{Zackrisson:2011aa} \citep[see also][]{Bromm:2001aa}.  

The above criteria reduces the number of possible galaxies at any redshift to $<10$. We make the final selection of fiducial galaxies by selecting the galaxy at each redshift with the smallest mean age of \piii star particles. We present the properties of these selected galaxies in Table~\ref{table:gals}. We present two values for the Star Formation Rate $({\rm SFR})$: Lifetime $\langle \sfr \rangle$ and Final $\sfr$. The former is the mean stellar mass formed per year since the galaxy started forming stars (i.e. the galaxy's total stellar mass divided by the age of its oldest star) while the later provides the average over the last $10\Myr$ of simulation run time (i.e. the stellar mass formed in the galaxy over the last $10\Myr$ divided by $10\Myr$). As stated in \S\ref{sect:intro} the majority of the \heii emission line is expected to be produced by massive stars which will have lifetimes on the order of $1$-$10\Myr$. We therefore adopt the upper age limit, i.e. $10\Myr$, as our time frame for calculating the Final $\sfr$. It is therefore expected that galaxies with a larger Final $\sfr$ will be brighter in \Heii. A comparison of the two $\sfr$ values can provide insight into the current state of star formation in each galaxy, e.g. if the Lifetime $\langle \sfr \rangle$ is greater that the Final $\sfr$ it suggests the galaxy is in a quiescent star formation phase.

All of the selected galaxies have $100\%$ of their stellar mass in \piii stars. By calculating the  mean (mass weighted) metallicity of the stars $\left(\langle Z_{\star}\rangle\right)$ in each galaxy, we find that the stars in our sample tend to have a metallicity of $\lesssim 0.27Z_{\rm crit}$ and with a small standard deviation around the mean (see Table~\ref{table:gals}). Finally we note that \emph{all} stars in both G3 and G7 have $Z_{\star}=Z_{\rm init}$.

\begin{sidewaystable*}
	\centering
	\rule{0cm}{20cm}
	\parbox{0.95\textwidth}{	
		\caption{Properties of Fiducial Galaxies}
		\begin{tabular}[h]{c c c c c c c c c c c c c c c c c c c}	
			\hline \hline
			Label  &Redshift & Stellar Mass & Gas Mass & Lifetime $\langle \sfr \rangle$ & Final $\sfr$        & $\langle Z \rangle$ & $\Delta x_{z}$ & $M_{\rm \star,PopIII,tot}/M_{\star,tot}$ & $\langle t_{\rm age,PopIII}\rangle$ & $\langle Z_{\star}\rangle/Z_{\rm crit}$ & $\sigma_{Z_{\star}}/Z_{\rm crit}$& Grating \\
			           &              &    $10^{6}\Msol$     &   $10^{7}\Msol$   &               $\Msol\,\yr^{-1}$       &$\Msol\,\yr^{-1}$ &    $10^{-3}Z_{\odot}$         & $\pc$  & & $\Myr$&         &            &\\
			{\bf (1)} & {\bf (2)} & {\bf (3)} & {\bf (4)} & {\bf (5)} & {\bf (6)} & {\bf (7)} & {\bf (8)} & {\bf (9)} & {\bf (10)} & {\bf (11)} & {\bf (12)} & {\bf (13)} \\
			\hline
			G1 & $9.98$   & $  7.1$ & $  8.5$ & $    0.073$ & $0.67$    & 4.4 & $24.6$  &  $1.0$ & $2.5$    &  $0.24$  &  $0.03$   & H+K \\  
			G2 & $8.98$   & $0.64$ & $  3.2$ & $    0.011$ & $0.056$  & 2.3 & $27.1$  &  $1.0$ & $0.91$  &  $0.15$  &  $0.03$   & H+K \\
			G3 & $8.04$   & $  1.5$ & $12.2$ & $     0.29$ & $0.15$     & 1.1 & $29.9$  &  $1.0$ & $0.57$  &  $0.05$  &  $0.00$   &H+K \\
			G4 & $6.97$   & $0.67$ & $  3.7$ & $   0.042$ & $0.046$   & 3.8 & $33.9$  &  $1.0$ & $2.95$  &  $0.26$  &  $0.07$   & Iz+J \\
			G5 & $6.02$   & $  1.7$ & $14.1$ & $   0.015$ & $0.14$     & 3.3 & $38.5$  &  $1.0$ & $1.5. $  &  $0.14$  &  $0.03$   & Iz+J \\
			G6 & $4.99$   & $  3.4$ & $39.9$ & $     0.11$ & $0.32$     & 2.8 & $22.6$  &  $1.0$ & $3.0$    &  $0.10$  &  $0.03$   & Iz+J \\
			G7 & $4.00$   & $60.7$ & $69.0$ & $    12.9$  & $6.1$       &0.99& $27.0$  & $1.0$  & $0.71$  &  $0.05$  &  $0.00$   & Iz+J\\
			G8 & $3.00$   & $  2.8$ & $29.7$ & $ 0.0026$ & $0.071$   & 6.5 & $33.8$  & $1.0$ & $70.0$   &  $0.27$  &  $0.12$   & V+R\\
	
			\hline
			\hline
		\end{tabular}\\
		{\footnotesize
		Notes: {\bf Column 1:} galaxy label, {\bf Column 2:} redshift of galaxy, {\bf Column 3:} galaxy's stellar mass, {\bf Column 4:} galaxy's gas mass, {\bf Column 5:} galaxy's $\sfr$ averaged over the age of the galaxy, {\bf Column 6:} $\sfr$ of the galaxy in the $10\Myr$ preceding ``observation'', {\bf Column 7:} mean metallicity of each galaxy, {\bf Column 8:} size of a single, fully refined cell at each redshift, {\bf Column 9:} Fraction of stellar mass in \piii stars, {\bf Column 10:} mean age of \piii stars, {\bf Column 11:} mass weighted metallicity of stars, {\bf Column 12:} Standard deviation of stellar metallicities, {\bf Column 13:} HARMONI grating to be used when observing galaxy 
		}
		\label{table:gals}
		}
\end{sidewaystable*}

\subsection{Modelling Stellar SEDs}
\label{meth:seds}
\begin{figure}
	\begin{center}
		\includegraphics[width=0.35\textwidth]{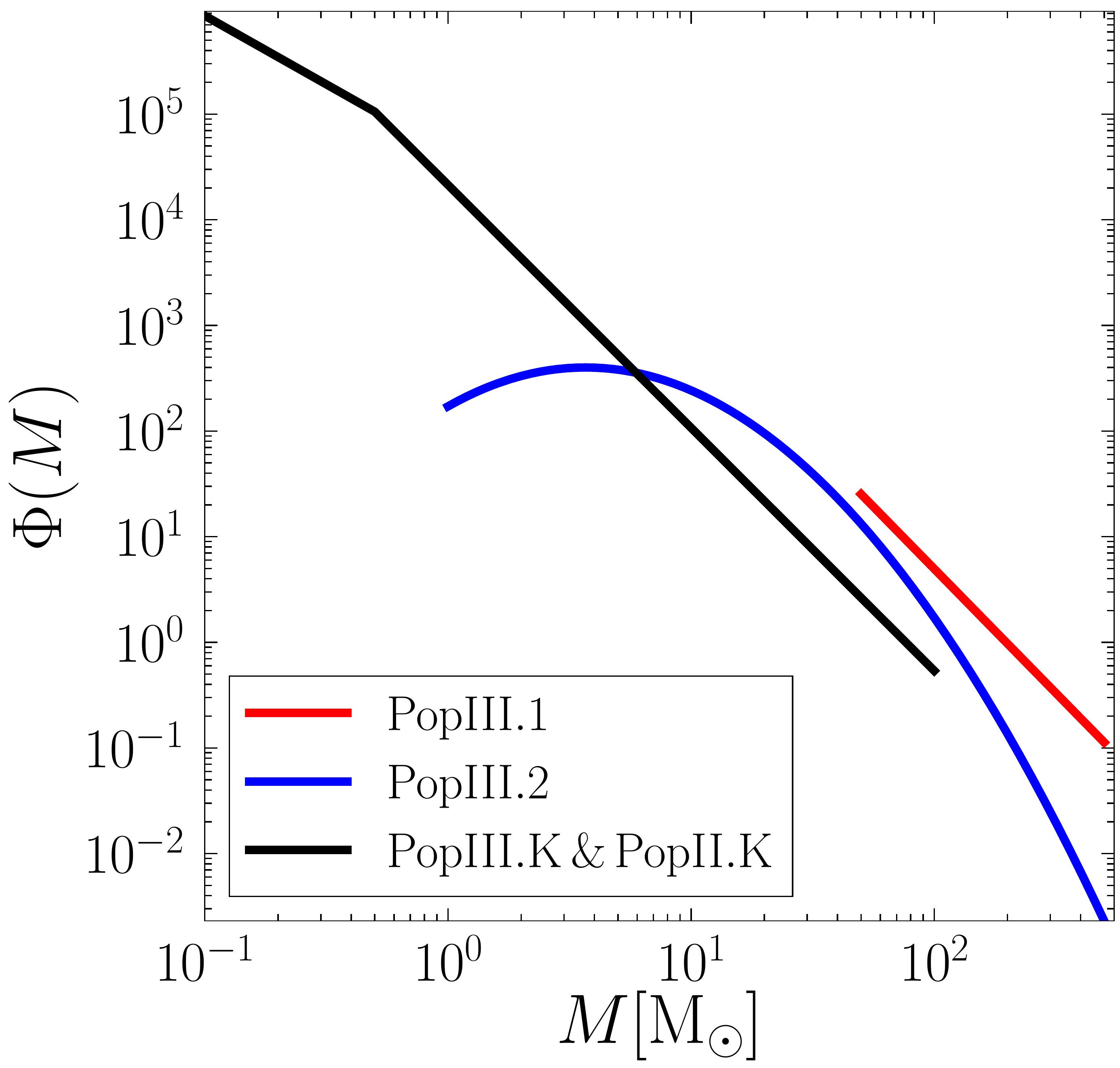}
		\caption{Initial Mass Functions (IMFs) for the three different SED models used throughout this work, see \S\ref{meth:seds} for details. Each IMF has been normalised so that the total mass of the stellar population is  $10^{5}\Msol$.		}
		\label{fig:imf}
	\end{center}
\end{figure}
As \newhorizon does not include explicit radiative transport, it is therefore necessary to post process the simulation to determine the spectrum of a given galaxy. This approach prevents us from drawing conclusions about how \piii stars will shape their environment and effect galaxy evolution. However, post processing does present an advantage: the choice of model used for \piii stars can easily be changed without rerunning the simulation. 

\begin{figure}
	\begin{center}
		\includegraphics[width=0.49\textwidth]{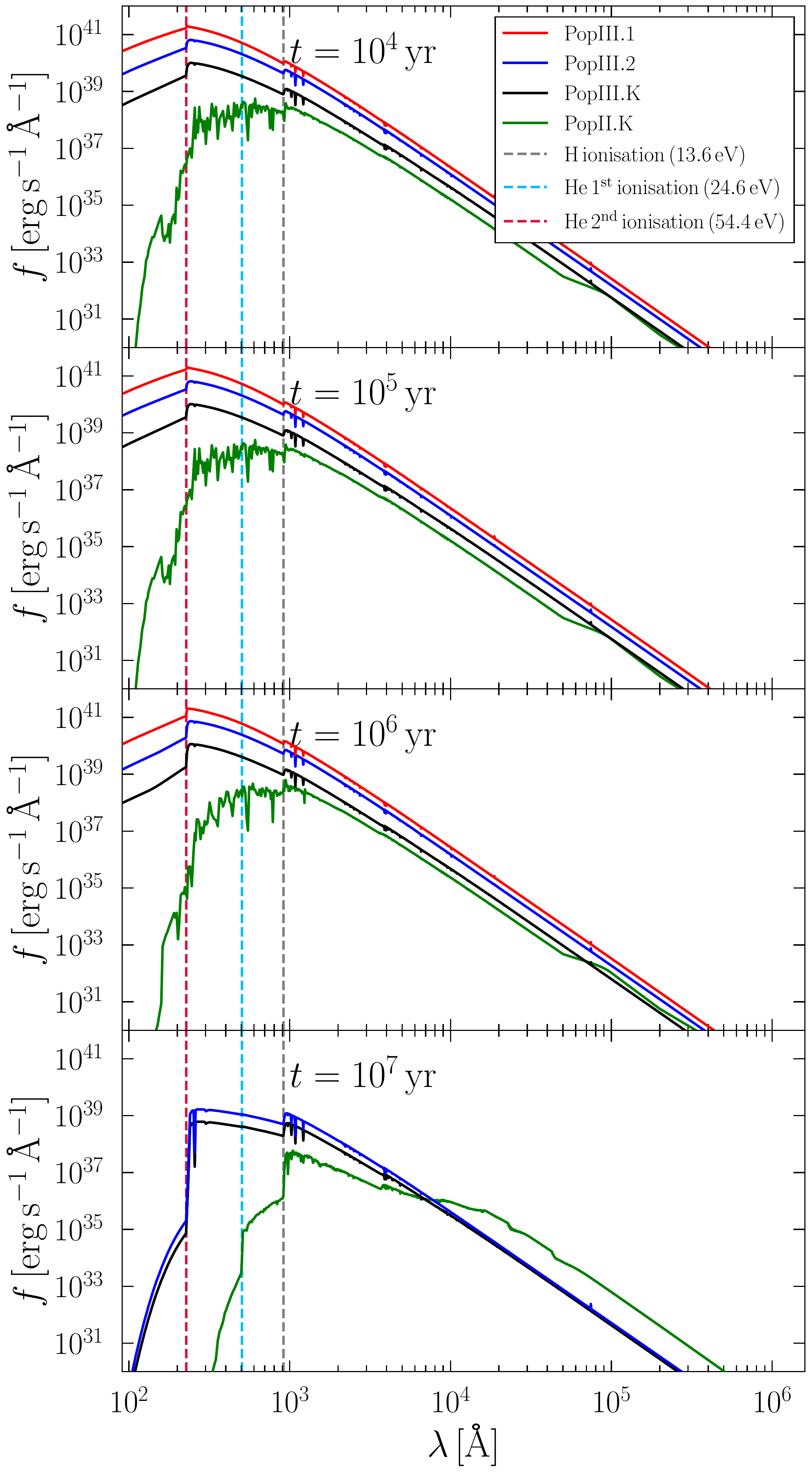}
		\caption{SEDs for the PopIII.1, PopIII.2,  PopIII.K and PopII.K IMFs (red, blue, black, and green solid lines respectively). Each panel shows the SEDs at a different age of the stellar population: $t=10^{4},\,10^{5},\,10^{6}$ and $10^{7}\yr$. The vertical dashed lines show the wavelengths at which He$^{+}$, He and H are ionised (red, blue and grey respectively). Due to the top-heavy IMF, the SED for PopIII.1 becomes zero for $t>3.6\times10^{6}\Myr$. All SEDs are for a stellar population with a total mass of $10^{5}\Msol$
		}
		\label{fig:sed}
	\end{center}
\end{figure}

The Spectral Energy Distribution (SED) of a population of stars depends on its IMF. The shape of the IMF at high $z$ is not yet constrained \citep[][]{Karlsson:2013aa} and is therefore a free parameter to be explored. In this work we employ three IMFs for \piii stars.  
The first (PopIII.1) assumes that the stellar population has an extremely top-heavy IMF, given by 
\begin{equation}
	\Phi(M) \propto \left\{
	\begin{array}{l l} 
		M^{-\alpha} &\quad\mbox{if $50\leq M\leq500\Msol$,}\\
 		0,\\
	\end{array}\right.
	\label{eq:imf1}
\end{equation}
where $\alpha=2.35$ \citep[see][]{Salpeter:1955aa,Schaerer:2002aa}. The second (PopIII.2) assumes a less extreme top-heavy IMF with a log-normal shape which is described by a characteristic mass of $M_{\rm c}=10\Msol$, a dispersion of $\sigma=1\Msol$ and wings extending from $1$ to $500\Msol$ \citep[see][for details]{Tumlinson:2006aa,Raiter:2010aa}. The final IMF (PopIII.K) uses the universal Kroupa \citep{Kroupa:2001aa} IMF, i.e.
\begin{equation}
	\Phi(M) \propto \left\{
	\begin{array}{l l} 
		M^{-0.3} &\quad\mbox{if $M<0.08\Msol$,}\\	
		M^{-1.3} &\quad\mbox{if $0.08\leq M<0.5\Msol$,}\\
 		M^{-2.3} &\quad\mbox{if $0.5\leq M\leq100\Msol$,}\\
		0.
	\end{array}\right.
	\label{eq:imf2}
\end{equation}
All three IMFs are shown in Fig.~\ref{fig:imf}, normalised for a $10^{5}\Msol$ stellar population. We use the {\sc yggdrasil} spectral synthesis code \citep[see][for details]{Zackrisson:2011aa} with the above IMFs to generate SEDs. The resulting SEDs, at four different stellar ages, are shown in Fig.~\ref{fig:sed}. All three models assume that the stellar population creating the SED has zero metallicity. 

\begin{figure}
	\begin{center}
		\includegraphics[width=0.45\textwidth]{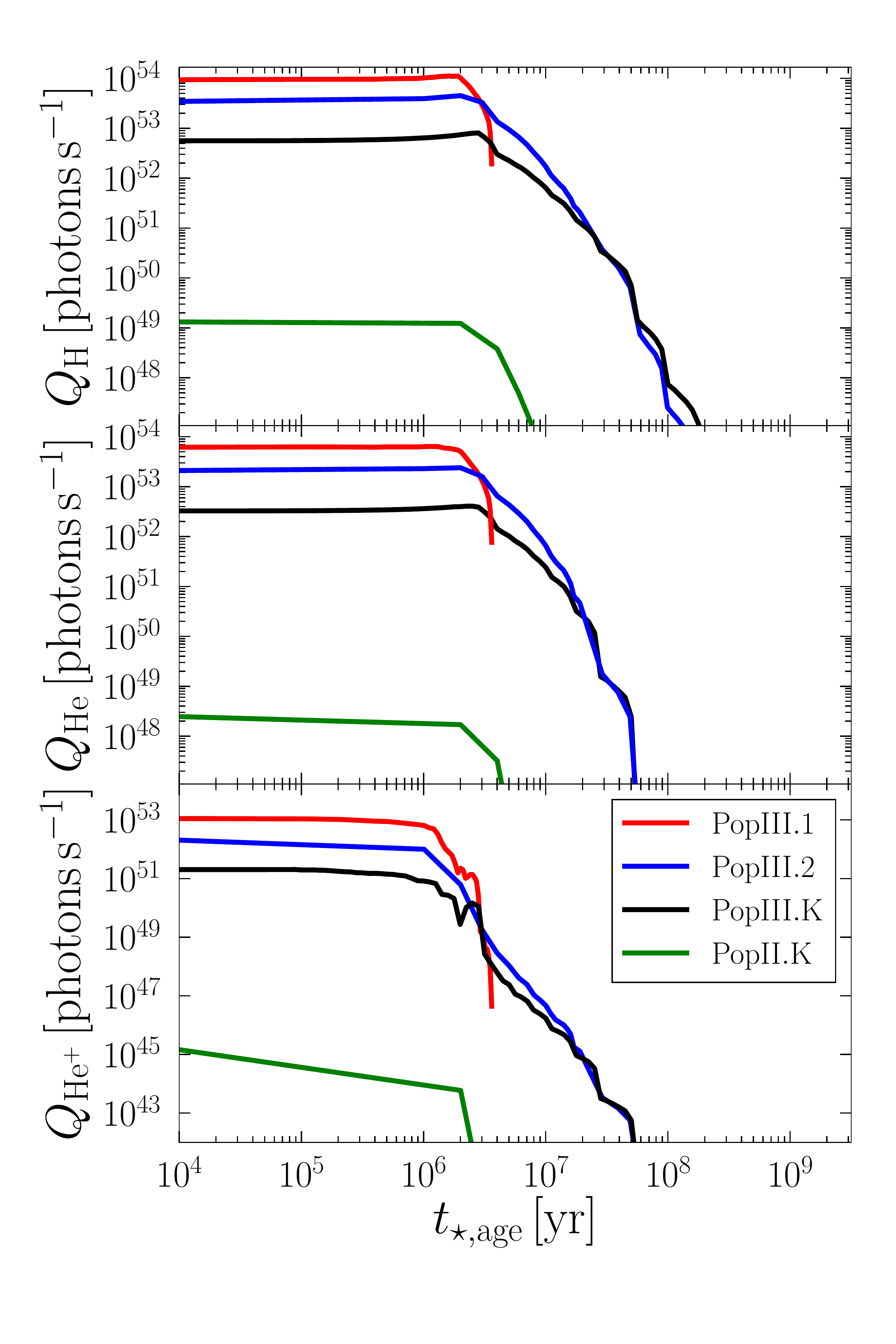}
		\caption{Total number of: H (top), He (middle) and He$^{+}$ (bottom) ionising photons per second ($Q_{\rm H},\,Q_{\rm He}$ and $Q_{\rm He^{+}}$ respectively) produced by each SED model as function of stellar age ($t_{\star,\rm age}$). In all 3 panels, red, blue, black and green lines show the evolution of $Q$ for ${\rm PopIII.1},\,{\rm PopIII.2},\,{\rm PopIII.K}$ and ${\rm PopII.K}$ SEDs. A stellar population with a total mass of $10^{5}\Msol$ was assumed when calculating $Q$. NB: The y-axis of each panel is scaled to best show the difference between all fours IMFs.  As a result of this choice, it appears that $Q_{\rm H }$ for ${\rm PopIII.1}$ is parallel to ${\rm PopIII.2}$ and ${\rm PopIII.K}$  or that $Q_{\rm H}/Q_{\rm He}$ or $Q_{\rm H}/Q_{\rm He+}$ is the same for each choice of IMF, however this is not the case. For example the latter ratio can vary by factor as large as $\sim3.5$ when comparing the different IMFs.
		}	
		\label{fig:Q}
	\end{center}
\end{figure}

As the stellar population ages the SED evolves and the number of photons produced per second with sufficient energy to ionise H, He, and He$^{+}$ ($E\geq13.6,\,24.6$ and $54.5\ev$ respectively) decreases, shown in Fig.~\ref{fig:Q}. All three of these models produces a large number of photons with energies $>54.5\ev$ at most ages. The exception being the PopIII.1 SED, which due to the extremely top-heavy nature of the IMF produces stars with short lifetimes ($<10^{7}\yr$). Despite this we use PopIII.1 as our fiducial model for \piii stars as it is the model most likely to be observable.  \newhorizon does not form individual stars but instead forms star particles which represent a cluster of gravitationally bound stars. We therefore assume that each star particle with $Z\leq Z_{\rm crit}$ represents a stellar cluster containing only \piii stars and we discuss the validity of this assumption in \S\ref{dis:valid}. 

As stated above, our selection criteria leads to all eight of our selected galaxies containing only \piii stars. However in principle it is possible that we could have selected galaxies that contain  non-\piii stars. In such a case additional \cloudy simulations would be needed, which account of each particle's metallicity and a non-\pii IMF.  Fig.~\ref{fig:sed} and \ref{fig:Q} (labelled as PopII.K) show an example of a SED and number of ionising photons produced by a stellar population following a Kroupa IMF with a metallicity of $Z_{\star}=0.02(=Z_{\odot}$).

The IMF used by the simulation at run time sets the amount of momentum, gas and metals injected into the surrounding ISM via feedback and thus determines galactic evolution. As stated above, \newhorizon employs a Chabrier IMF and as a result of this work imposing a different IMFs in post-processing the self-consistency of the simulation is lost. We can not, therefore, use this simulation to explore how \piii stars following the ${\rm PopIII.1}$, ${\rm PopIII.2}$ or ${\rm PopIII.K}$ IMFs impact a galaxy's structure, dynamics or evolution. However, as their is no radiative transfer in \newhorizon and adding photons to the simulation in post processing does not change the gas dynamics we are still able to make predictions about the observability of \piii stars.  
 

\subsection{Constructing Spectra}
\label{meth:spec}
For each galaxy we extract a cube, with sides of $4\kpc$, centred on the galaxy, this volume is divided into a uniform grid of cells with sides of $\Delta x_{z}$ (see Table~\ref{table:gals} for values at a given $z$). Each cell has its own unique gas properties (e.g. density,  temperature and velocity) and associated star particles. To create an ``observable'' Spatial-Spectrum Data Cube (SSDC) of a galaxy we construct a spectrum for each cell (and hence star particles). Below we outline the method of its construction. 

\subsubsection{\cloudy Models}
\label{meth:cloudy}
The \heii emission line is not produced directly by \piii stars but instead is the result of their radiation being reprocessed by the surrounding gas. To model these processes we employ the microphysics code \cloudy \citep[see][for full details on \cloudyp]{Ferland:2017aa}. We assume that each star particle is at the centre of a spherical gaseous cloud (of constant density) with a radius $\Delta x_{z}/2$. \cloudy is used to calculate the strength of emissions lines and the continuum at the outer edge of the cloud, i.e. out to a radius of $\Delta x_{z}/2$. In addition to the radiation field from stars we include a background radiation field. This field is designed to mimic the observed cosmic radio to X-ray background with contributions from the CMB, assumed to be a black body with a temperature of $T_{\rm CMB}=2.725(1+z){\rm K}$. To explore all possible environments surrounding star particles we run a grid of \cloudy models, at each redshift, $z$, varying: gas number density ($n$), stellar age ($t_{\star,\rm age}$) and total luminosity of SED ($L_{\rm tot}$). A full sweep of parameter space is completed for all three sets of SEDs (PopIII.1, PopIII.2 and PopIII.K) so that the impact of different IMFs can be explored. 

All  \cloudy models with \piii IMFs use the primordial element abundances i.e. $76\%$ of the gas (by mass) is hydrogen, with the remainder made up of Helium. As a result we set \cloudy to only record line intensities for H and He lines in the wavelength range of interest ($1000\leq\lambda\leq6600\ang$), i.e. we assume that all \piii stars form from pristine gas and do not pollute their environment until they die. 

There are a number of cells both in and surrounding our selected galaxies that contain no star particles. For these cells we run additional \cloudy models using the background radiation field but without a radiation field from the stars. As a result the only parameters varied in these \cloudy simulations at a given $z$ is $n$, as we let \cloudy determine the equilibrium temperature. 

We have compared the \cloudy method described above with a simple analytical model (see Appendix \ref{app:ana}) and find good agreement.

\subsubsection{Constructing Spectra for Cells with Stars}
\label{meth:stars}
If a cell has one or more star particle, a spectrum is calculated for each particle which are then summed to create a single spectrum for that cell. We create the spectrum for each star particle as follows. First, we select the SED which matches both the star particle's age ($t_{\star,\rm age}$) and the IMF of choice. 
When creating SEDs, {\sc yggdrasil} assumes the birth mass, $M_{\rm SED}$, of the stellar population which is combined with IMF to determine the shape and magnitude of the SED (i.e. the number of photons at each wavelength). Each star particles in \newhorizon has a birth mass, $M_{\star,\rm birth}$, which is rarely equal to $M_{\rm SED}$. It is therefore necessary to scale the magnitude of the SED from {\sc yggdrasil} so that the correct value of $L_{\rm tot}$ can be passed to \cloudyp. This is calculated using 
\begin{equation}
	L_{\rm tot} = \frac{M_{\star,\rm birth}}{M_{\rm SED}}\int L_{\lambda}{\rm d}\lambda,
	\label{eq:Ltot}
\end{equation}
where $L_{\lambda}$ is luminosity of the SED at each wavelength  ($\lambda$). 
$M_{\star,\rm birth}$  therefore scales the magnitude of the SED while the choice of IMF and $t_{\star,\rm age}$ determine its shape for each particle. 

{\sc yggdrasil} only provides SEDs until the  time at which the first supernova is expected for each SED (e.g. $t_{\rm SN}=10^{6.5},\,10^{7.5}$ and $10^{8}\Myr$ for ${\rm PopIII.1,\,PopIII.2}$ and ${\rm PopIII.K}$ respectively) therefore if $t_{\star,\rm age}>t_{\rm SN}$ the star particle is marked as ``dead'' and does not contribute to the spectrum of its host cell. In principle ``dead'' particles will still contribute to luminosity of their host galaxy, either as cooling supernova remnants or as low mass ($\sim1\Msol$) Main Sequence stars in the case of ${\rm PopIII.K}$. As this will not effect the \heii line, on which this work is focused,  we have chosen not to include this contribution. 

With $t_{\star,\rm age}$ and $L_{\rm tot}$ of the particle as well as the $n$ of its host cell and the determination of \pii or \piii the particle is matched to an appropriate \cloudy model. We take each line's intensity ($I_{\rm 0}$) from \cloudy and assume that each line is a Gaussian, given by
\begin{equation}
	I(\lambda) = I_{\rm norm}e^{-\frac{(\lambda-\lambda_{\rm 0})^{2}}{2\sigma^{2}}},
	\label{eq:gaus}
\end{equation}
where $\lambda_{\rm 0}$ is the rest wavelength of the line, $\sigma$ sets the width of the function and $I_{\rm norm}$ is a normalisation constant to ensure that $\int I(\lambda)\diff\lambda=I_{\rm 0}$. We assume that the Full Width Half Maximum (FWHM) of the Gaussian is set by the thermal motions of the gas, i.e.: FWHM$\,= 2\sqrt{2\ln2}\Delta v_{\rm g,therm}$ where 
\begin{equation}
	\Delta v_{\rm g,therm} = \lambda_{\rm 0}\sqrt{\frac{k_{B}T_{g}}{m_{a} c^{2}}}=\sigma,
\label{eq:gasbroad}
\end{equation}
here $T_{g}$ is the gas temperature in the cell, $m_{a}$ is the mass of the element (hydrogen or helium) emitting the line, $c$ is the speed of light and $k_{B}$ is the Boltzmann constant \citep{Blundell:2006aa}. 
 
Once the shape and intensity of each emission line is known, it is added to the continuum emission (also given by \cloudyp) to create the entire spectrum for the particle. 
Finally the spectra from all star particles in a given cell are summed to give the total spectrum for that cell.

\subsubsection{Constructing Spectra for Cells Without Stars}
Cells that have been identified as not containing a star particle are matched to an appropriate \cloudy model (i.e. those only employing a background radiation field, see \S\ref{meth:cloudy}) using just $n$. As with cells containing star particles, the spectrum of a cell without stars is constructed using line and continuum intensities from \cloudyp.

\subsubsection{Constructing Data Cubes}

Next we modify the spectrum for each cell based on the turbulent motions of the gas within. This is achieved by convolving the spectrum of a cell with a Gaussian whose width is given by the 3D velocity dispersion of the gas ($\sigma_{\rm g,vel}$) i.e. the turbulent motions. We define $\sigma_{\rm g,vel}$ as $\sqrt{(\sigma_{g,x}^{2}+\sigma_{g,y}^{2}+\sigma_{g,z}^{2})/3}$, where $\sigma_{g,x},\,\sigma_{g,y}$ and $\sigma_{g,z}$ are the velocity dispersions of the three velocity components of a cell. 

Next wavelength-dependant extinction, due to dust, along the line of sight is added to each cell's spectrum. We achieve this by first calculating $A_{V}$ using 
\begin{equation}
	A_{V}=1.086\frac{3f_{\rm d}\Sigma_{\rm Z}Q_{\lambda}}{4\rho_{\rm d}\sqrt{a_{1}a_{2}}},
\end{equation}
where $Q_{\lambda}=1.5$ is a constant extinction coefficient, $\rho_{\rm d}=4{\rm \,g\,cm^{-3}}$ is the typical density of a dust particle, $f_{\rm d}=0.1$ is the fraction of gas-phase metals locked up in dust, $\Sigma _{\rm Z}$ is the column density of metals, $a_{1}=0.005{\rm \,\mu m}$ is the smallest size of a dust grain and $a_{2}=1{\rm \,\mu m}$ is the largest size of a dust grain \citep[see][for full details of the method and discussion on choice of values]{Richardson:2020aa}. As $\Sigma _{\rm Z}$ depends on the mass of metals along the line of sight between a given cell and the observer, a unique value is calculated and used for each cell. We combine $A_{V}$ with the dust extinction curve ($E(\lambda)$) found by \cite{Fitzpatrick:1999aa}, assuming $E(\lambda) = A_{\lambda}/A_{V}$, here $A_{\lambda}$ is the wavelength specific extinction. Thus intensities of each cell spectrum is given by
\begin{equation}
	I_{\rm E}(\lambda)=I(\lambda)10^{\frac{A_{V}E(\lambda)}  {-2.5}},
\end{equation}
here $I(\lambda)$ is the spectrum of a cell before extinction is applied and $I_{\rm E}(\lambda)$ is the spectrum after extinction is applied. We do not include the extinction due to the gas within the cell we are currently considering as this is applied by \cloudyp. As a result of the dependance on metallicity and that our choice galaxies have very low metallicity (see Table\,\ref{table:gals}), both $f_{\rm d}$ and $\Sigma_{\rm Z}\sim0$ for most cells and thus the impact of extinction is negligible, however it is included for completeness. The above is a simple model of extinction and does not account for every process that is able to reduce the strength of an emission line. 
In order to fully capture every possible process a new series of simulations with full radiative transfer would be required where the path of each photon is taken into account \citep[e.g.][]{Laursen:2007aa} which is beyond the scope of the work presented here.

Before the spectra along the line of sight are combined we Doppler shift the spectrum based on the gas line of sight velocity ($v_{\rm g,los}$) using 
\begin{equation}
	\lambda_{\rm o}=\lambda_{\rm e}(\frac{v_{\rm g, los}}{c}+1),
	\label{eq:lshift}
\end{equation}
where $\lambda_{\rm o}$ is the observed wavelength and $\lambda_{\rm e}$ is the emitted wavelength. This shift in wavelength should not be confused with the shift cause by placing each galaxy at the appropriate redshift, i.e. at this stage we treat the galaxy, its stellar and gaseous contents as if it is at $z=0$. We then sum the intensities at each wavelength along the line of sight to create an SSDC of the simulated galaxy. Finally, we place the galaxy at the appropriate redshift by shifting the wavelength range (e.g. for G1 $1640\ang\rightarrow18040\ang$) and convert the luminosity of each spaxel into flux by accounting for the luminosity distance (i.e. dividing by $4\pi D_{\rm L}^{2}$).

\subsection{Observing the Simulations: \hsim}
\label{meth:hsim}
We employ the HARMONI simulator \hsim (version 2.10) to produce mock \eelt observations of each SSDC.  \hsim is a {\em cube-in, cube-out} simulator for HARMONI that replicates all known instrumental effects, including the strongly wavelength-dependent adaptive optics (AO) point spread function (PSF) \citep{Zieleniewski:2015aa}. It also includes thermal background from telescope and instrument, as well as night-sky emission lines and continuum (include lunar contributions), and telluric absorption. Instrumental transmission, detector read noise, detector cross-talk, shot noise from all background sources is also modelled.  \hsim also incorporates the instrumental wavelength response (line spread function, or LSF), and the impact of spatial sampling for the chosen spaxel scale.  The resulting data cube in {\tt FITS format} is in units of ${\rm electrons\,s^{-1}}$. An associated variance cube is also produced. These can be used to analyse mock observations as if observed with HARMONI at the \Eelt.  \hsim v2 is a completely re-coded version that adopts a {\em follow-the-photons} philosophy for the simulations, adding noise and background flux, and applying throughput losses in the same order as they occur along the sky, telescope and instrument light path \citep{Pereira-Santaella:2019aa}.

\begin{table}

		\parbox{0.5\textwidth}{	
			\caption{{\sc hsim Settings}}
			\begin{tabular}[h]{l r }		
				\hline \hline
				Parameter  & Settings  \\
				                   &                \\
				\hline
				Exposure Time (s)                           &    900    \\  
				Number of Exposures.                     &    40      \\
				Spatial Pixel Scale (mas), $z\geq6$&$10$x$10$\\
				Spatial Pixel Scale (mas), $z<6$&   $20$x$20$\\
				Adaptive Optics Mode                      &    LTAO \\
				Zenith seeing ('')                               &    0.43   \\
				Air Mass                                           &    1.3  \\
				Moon Illumination                             &    0.0     \\
				Telescope Jitter sigma (mas)           &    3.0     \\
				Telescope Temperature (K)              &    280    \\
				Atmospheric Differential Refraction  &    True  \\
				Noise Seed                                       &    1.0     \\
				\hline
				\hline
			\end{tabular}\\
			\label{table:hsim}
		}	

\end{table}

\begin{figure*}
	\begin{center}
		\includegraphics[width=1.0\textwidth]{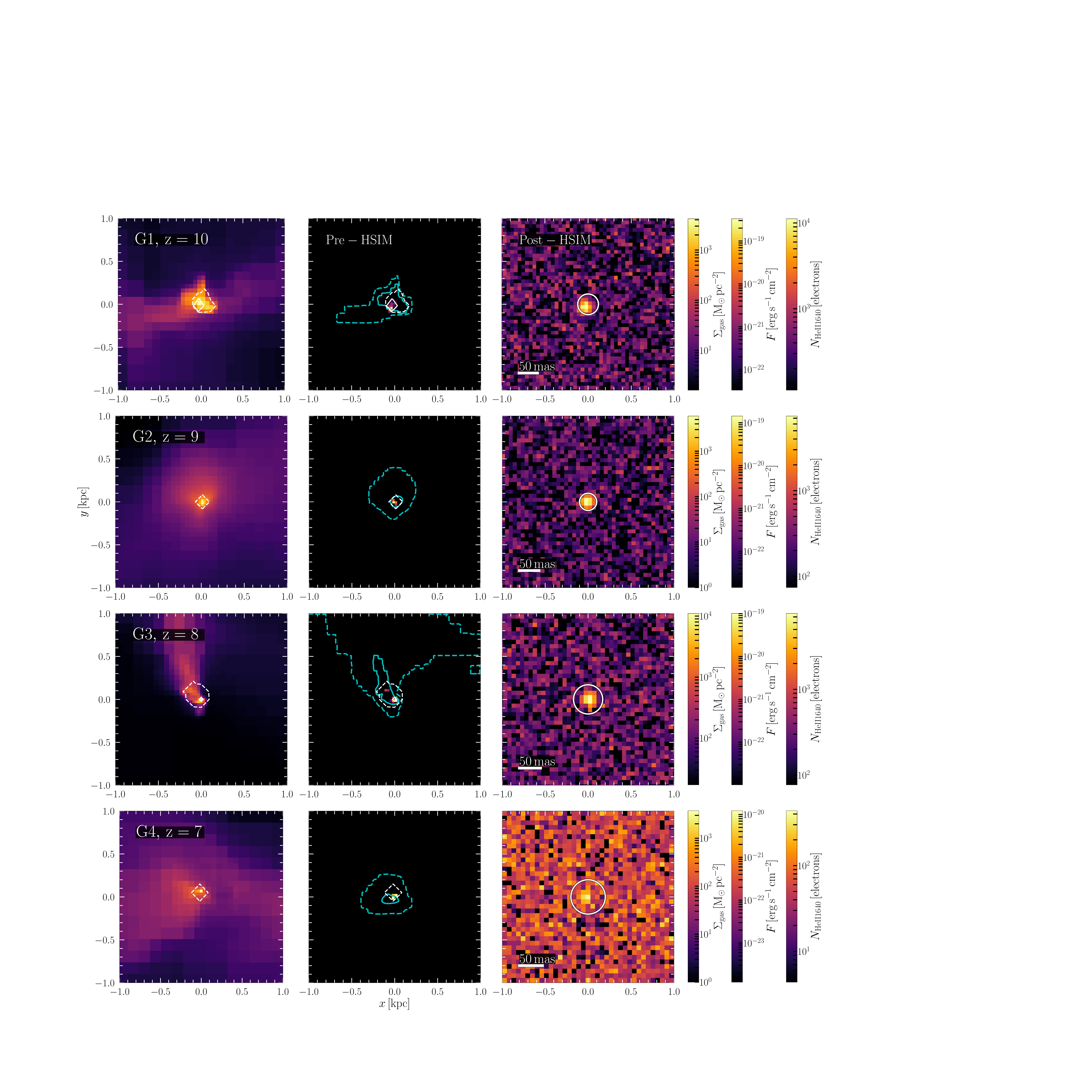}
		\caption{Left: Gas surface density ($\Sigma_{\rm gas}$) map. Middle: \heii Integrated line strength map before observation with \hsimp. Right: \heii Integrated line strength map after observation with \hsimp. The white contours show regions of the galaxy which have stellar surface densities of $\Sigma_{\star}\geq1$ and $100\Msol\,\pc^{-2}$ (dashed and solid contours respectively). The cyan contours (middle \& right) shows $\Sigma_{\rm gas}$ of $\geq10^{1.5}$ and $ 10^{2.5}\Msol\pc^{-2}$ (dashed and solid contours respectively). The aperture used for spectrum extraction in \S\ref{res:sas} are shown by the white circle shown in the right column. Each row shows the maps for one of the galaxies described in Table~\ref{table:gals}. The colour scales for each row is given on the far right, in the same order as the maps. The white bar in the bottom-left of each of the right hand side panels shows the size of a $50 \,{\rm mas}$ region at a redshift of each galaxy. Due to the switch to the $20\mas$ spaxels for galaxies at $z<6$ when observing with \hsim the there is a visible increase in size of the map pixels for G6, G7 and G8 in the right hand column. NB: only the ${\rm PopIII.1}$ IMF is shown in this figure. 		}
	\end{center}
\end{figure*}
\begin{figure*}
	\ContinuedFloat
	\addtocounter{figure}{0}
	\begin{center}
		\includegraphics[width=1.0\textwidth]{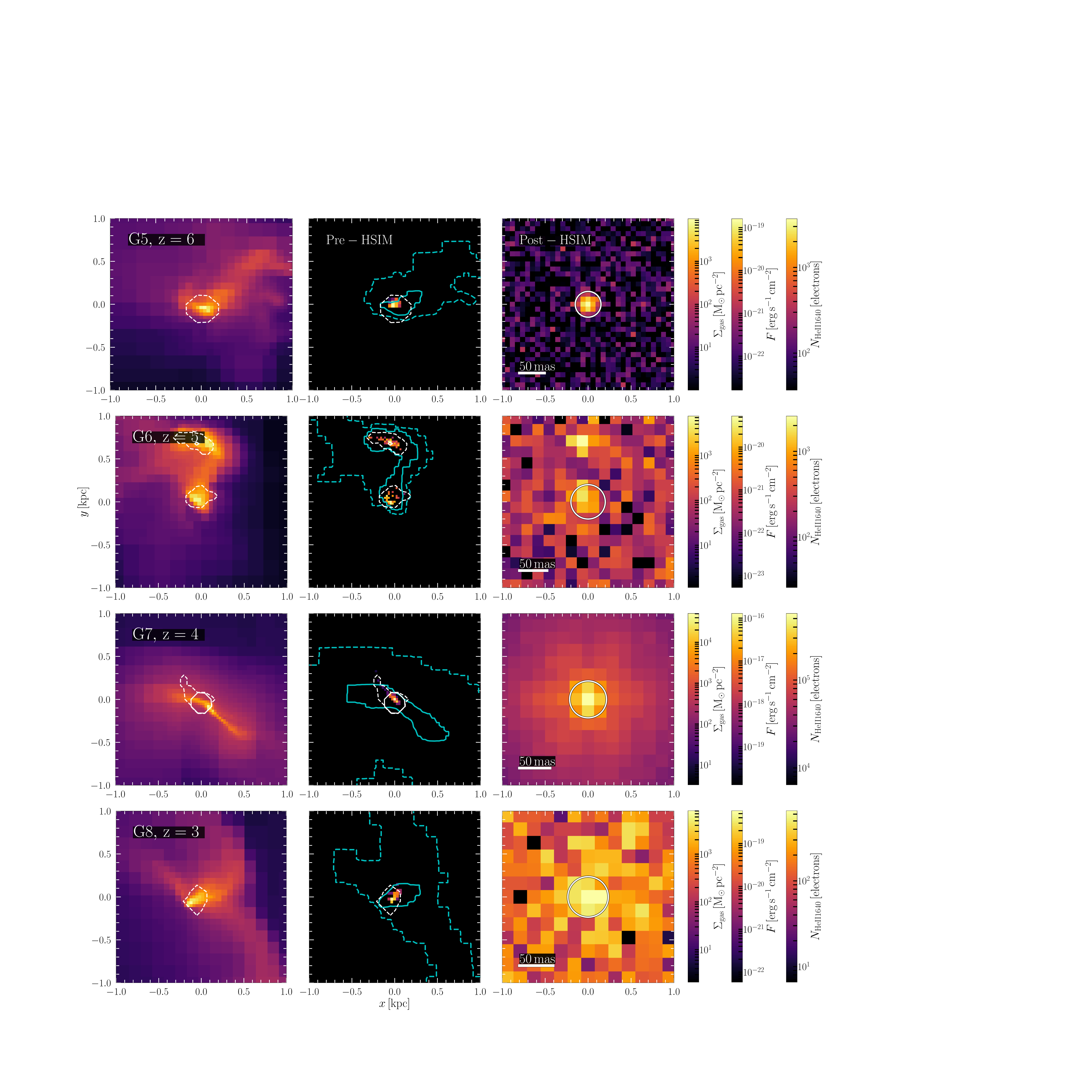}
		\caption{Continued
		}
		\label{fig:maps}
	\end{center}
\end{figure*}

For our analysis we employ the \hsim settings outlined in Table \ref{table:hsim} (unless otherwise stated). These settings are chosen to provide the most sensitive observation, therefore the results presented in this work should be considered the \emph{best case}.  As the choice of grating is determined by the redshift of the observed object we give the grating used for each galaxy in Table~\ref{table:gals}. We explored using the $10$x$10\mas$ spaxels at all $z$ but found that at $z<6$ the \heii signal was lost in the noise.  We have therefore opted to use  the $20$x$20\mas$ spaxels for $z<6$ as these are more sensitive than the $10$x$10\mas$ spaxels.  

\section{Results}
\label{sect:results}

\subsection{\newhorizon Galaxies at $1640\ang$}
\label{res:gen}

Due to our selection criteria, the galaxies in our sample all consist of a central high density region ($\Sigma_{\rm gas}\gtrsim10^{3}\Msol\pc^{-2}$), which has radius on the order of $100$ parsecs. Additionally most galaxies also show some larger scale extended, but lower density structures ($\Sigma_{\rm gas}\sim10^{2}\Msol\pc^{-2}$), surrounding the central region (see the left column of Fig.~\ref{fig:maps}). The stellar structure of the galaxies varies more than the gas structure (as shown by the contours in the left column of Fig.~\ref{fig:maps}). That being said we do find a trend that as $z$ decreases the stellar structures tends to become more extended.

\begin{figure*}
	\begin{center}
		\includegraphics[width=0.95\textwidth]{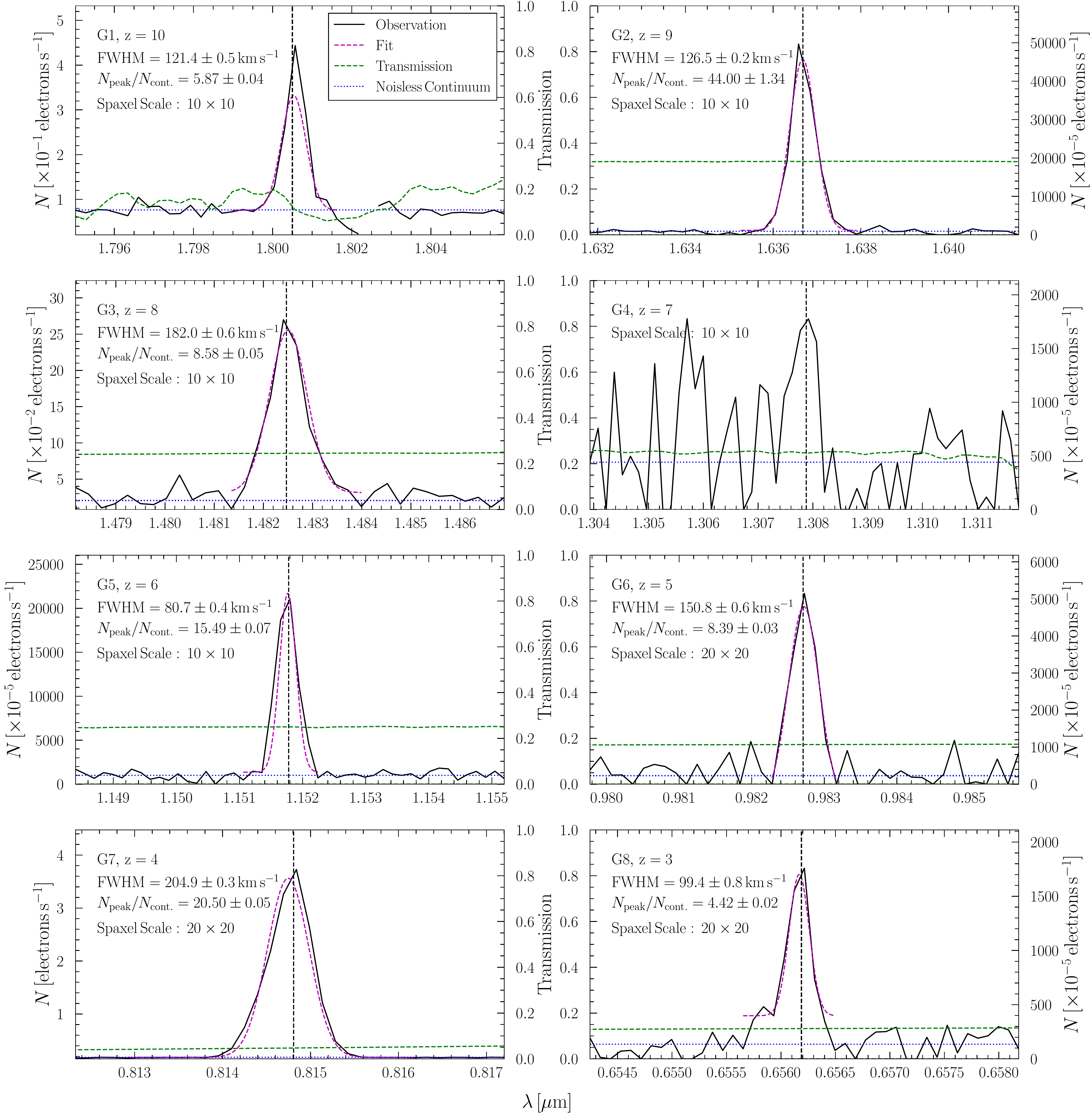}
		\caption{Single Aperture Spectra for the eight fiducial galaxies (solid black line). The dashed-black line indicates the wavelength of the peak emission line prior to observations with \hsimp. A fit to this emission line found in this spectrum is given by the magenta line, with the FWHM and peak to continuum ratio of the fit given in each panel. The green dashed line shows the fraction of light at a given wavelength reaching the detector. To demonstrate the impact of noise on \piii detectability we include the continuum extracted from the noiseless \hsim data cube (blue dashed line).}		
		\label{fig:lineprof}
	\end{center}
\end{figure*}

\subsubsection {Flux Maps}
\label{res:fmaps}
\begin{table}

		\parbox{0.4\textwidth}{	
			\caption{Galaxy Luminosities in \heii}
			\begin{tabular}[h]{c c c c}		
				\hline \hline
				Label  & &$L_{1640\lambda} [10^{42}\rm erg\,s^{-1}]$ &\\
				\cline{2-4}
				           & ${\rm PopIII.1}$ & ${\rm PopIII.2}$ & ${\rm PopIII.K}$ \\ 
				{\bf (1)} & {\bf (2)} & {\bf (3)} & {\bf (4)} \\
				\hline
				G1 &  $6.26$       &  $1.72$       &  $0.231$   \\  
				G2 &  $1.72$       &  $0.357$     &  $0.0411$ \\
				G3 &  $1.39$       &  $0.365$     &  $0.0530$ \\
				G4 &  $0.107$     &  $0.0485$   &  $0.0103$ \\
				G5 &  $1.17$       &  $0.543$     &  $0.0923$ \\
				G6 &  $0.502$     &  $0.124$     &  $0.0231$ \\
				G7 &  $170$        &  $37.8$       &  $5.26$     \\
				G8 &  $0.40$       &  $0.107$     &  $0.0210$ \\
				\hline
				\hline
			\end{tabular}\\
			{\footnotesize
			Notes: {\bf Column 1:} galaxy label, {\bf Column 2:} Total integrated Luminosity in the \heii emission line for the ${\rm PopIII.1}$ IMF, {\bf Column 3:} Total integrated Luminosity in the \heii emission line for the ${\rm PopIII.2}$ IMF, {\bf Column 4:} Total integrated Luminosity in the \heii emission line for the ${\rm PopIII.K}$ IMF.
			}
			\label{table:lumi}
		}	
\end{table}
Running each galaxy through our SSDC generation pipeline (using the ${\rm PopIII.1}$ IMF) we find that all eight galaxies produce a \heii emission line from which we are able to calculate the integrated \heii luminosity ($L_{1640\lambda}$) which are given in the second column of Table~\ref{table:lumi} and the middle column of Fig.~\ref{fig:maps} respectively. The line strength is always greatest in regions of high gas and stellar density. Furthermore, (reassuringly) the \heii emission is only found in pixels containing stars and not being produced by the locations with only gas. In \S\ref{meth:nh:sel} we stated the expectation that galaxies with larger Final $\sfr$ will also have larger $L_{1640\lambda}$, which in general we find to be the case by comparing columns 6 and 2 of Tables \ref{table:gals} and \ref{table:lumi} respectively. As a result of our selection criteria all of our galaxies are spatially very compact, especially in terms of their stellar structure. This results in the emission from these galaxies being found in only a handful of pixels. We have run our pipeline on more extended low redshift spiral galaxies and found that we are able to recover the full extent of these galaxies and their structure in flux maps.
 
After being ``observed'' with \hsim (right column of Fig.~\ref{fig:maps}) we find that the \heii line is detectable in seven of the eight galaxies: G1, G2, G3, G5, G6, G7 and G8. We are unable to detect any signal from G4 after observations with \hsimp, however given that galaxies at larger $z$ are detectable we conclude that the lack of signal from this galaxy is due to its stellar mass and age combined with gas mass and  morphology (see Table~\ref{table:gals}). These maps also show that, for the more extended galaxies, their morphology cannot be recovered.

It is worth noting that at $z=3$ the \heii emission line is redshifted to the observed V band, a wavelength regime where the Laser Tomography Adaptive Optics (LTAO) correction is very poor, with Strehl ratios typically below $1\%$.  While the PSF still has a small diffraction limited core, the fraction of the incident energy in the diffraction limited core is very small, with most of the light spread over a region of size $\sim200\,{\rm mas}$. Thus, only the strongest compact emission features from the galaxy are detected, resulting in the emission line only being seen in a single spaxel.

\subsubsection {Single Aperture Spectra}
\label{res:sas}

To each map we apply a circular mask centred on the galaxy (the mask is shown as a white circle in the right column of Fig.~\ref{fig:maps}). By summing the spectrum of each spaxel within the mask we produce a single-aperture spectrum for each galaxy, (Fig.~\ref{fig:lineprof}, black lines). Each galaxy uses a different size mask, where the size is set to maximise the signal to noise in resulting spectrum. In addition to the aperture we also calculate the variance for each spaxel at each wavelength and mask the signal from those with high variances ($\sim15\times$ that of the minimum calculated for each observation), typically those wavelengths affected by strong sky emission lines or strong telluric absorption features. 

For each spectrum we fit a Gaussian (magenta line in Fig.~\ref{fig:lineprof}) to the strongest emission feature (i.e. the  \heii emission line) and calculate its Full Width Half Maximum (FWHM) and the ratio of the emission line peak to continuum ($N_{\rm peak}/N_{\rm cont.}$). To ensure that the emission line recovered is in fact the \heii line, we mark its pre-\hsim position in each panel (black dashed line). We present the recovered values for each galaxy in their respective panel in  Fig.~\ref{fig:lineprof}, which shows that the \heii line is detectable in seven of eight galaxies. As before, we find that the signal from \piii stars in G4 is lost in the noise and thus would not be detected by HARMONI. 

Each panel of Fig.~\ref{fig:lineprof} shows the transmission curve (i.e. the fraction of light that reaches the detectors after traveling through the Earth's atmosphere, the \eelt and HARMONI) used by \hsimp. From these curves we see that detection of \heii at certain redshifts will be more susceptible to noise and instrument effects than others. For example only $\sim5\%$ of the \heii light emitted by galaxies at $z=4$, such as G7, will reach the detectors. To determine if the transmission curve is preventing a detection of G4, we artificially change the wavelength of the galaxy's spectrum to that of a galaxy at $z=7.5$\footnote{To ensure that we are only testing the transmission curve we keep G4 at a luminosity distance equal to that of $z=7$ galaxy.} and re-observe it with \hsim in the H+K band. Unfortunately, the \heii line from G4 is still undetectable and we therefore conclude that the intrinsic brightness of G4 is simply not sufficient for \piii stars to be detected.

\subsection{Impact of Luminosity Distance}
\label{res:id}

\begin{figure}
	\begin{center}
		\includegraphics[width=0.48\textwidth]{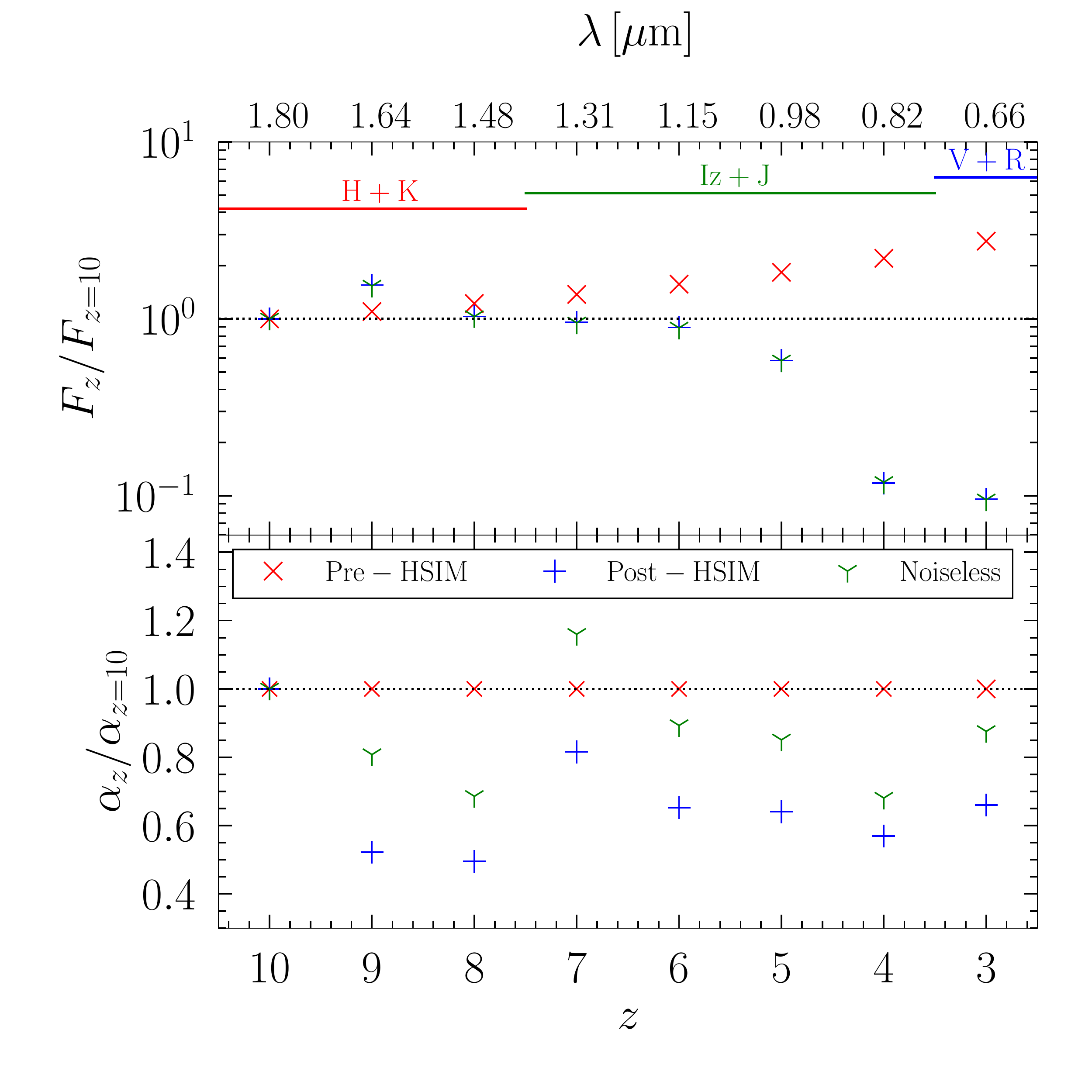}
		\caption{Top: Peak line strength of the \heii emission line ($F_{z}$) in G1 when the galaxy is moved to different redshifts ($z$). Bottom: Ratio of peak line strength to continuum ($\alpha_{z}$) when G1 is moved to different $z$.  Red ``x''  and blue ``+'' show the values measured before and after observation with \hsim respectively. Green ``Y'' gives the value of $F_{z}$ and $\alpha_{ z}$ when calculated from the \hsim data cube without noise. The values have been normalised to the value of $F_{z=10}$, i.e. the value at G1's original redshift. 
			}	
		\label{fig:dl}
	\end{center}
\end{figure}

In this work our ``observed'' galaxies have luminosity distances ($D_{\rm L}$) between  $\sim20$ and $\sim110\Gpc$ and while it is obvious that the luminosity of an object decreases as $D_{\rm L}$ increases, it is not clear how $D_{\rm L}$ will impact the observability of a given object. 
By artificially placing one of our galaxies (G1) at different redshifts we are able to determine the role of $D_{\rm L}$\footnote{Changing $D_{\rm L}$ also results in changes to the angular size of the galaxy.} in measuring the line strength. As expected, moving G1 from $z=10$ to $z=3$ results in an ($\sim3\times$) increase of the line flux ($F$) of the galaxy before observation while the ratio of the emission line peak to continuum ($\alpha$) remains constant (see Fig.~\ref{fig:dl}).

When observing G1 with \hsim after moving it to smaller $z$ we do not recover the systematic increase in $F$ found before observations (see Fig.~\ref{fig:dl}); instead $F$ tends to decrease with $z$. Furthermore, $\alpha$ is no longer constant, instead it varies with $z$.
This is the result of a combination of different factors: the post-\hsim line flux appears to decrease because the AO PSF has a substantially smaller fraction of the total energy enclosed in the PSF core as we move to shorter wavelengths (lower $z$). Thus, although the total line flux increases, the fraction of the flux within the extraction aperture drops even more markedly, resulting in a net decrease of line flux.  In addition, variations in spectrograph throughput as a function of wavelength (mostly stemming from the grating's efficiency curve) also play a role.  One can understand the variation of line to continuum ratio by considering the AO PSF as having two components: one has the size of the diffraction core of the \eelt, and the other is several hundred milli-arcseconds in size, close to the seeing disc. Depending on the structure of the galaxy's line and continuum emission, different fractions of each couple into the extraction aperture at different wavelengths, as the ratio of the two components of the PSF change dramatically, by more than an order of magnitude \citep[for full details we refer the reader to][]{Zieleniewski:2015aa,Pereira-Santaella:2019aa}.  

In short HARMONI (and thus the \Eelt) has a wavelength dependant instrumental response which produces a non-trivial dependance on $z$ to the sensitivity of line fluxes. 
A full analysis of these different processes impacting observations is beyond the scope of the current work and we leave it for future studies. However, assuming that  \hsim provides a reasonable representation of HARMONI, we conclude that the redshift of an object is not the determining factor in how bright that object will appear in observations. 

\subsection{Impact of IMF on Observability}
\label{res:imf}
The results in the previous sections (\S\ref{res:gen} \& \ref{res:id}) assumed an extremely top-heavy IMF for \piii stars when applying our pipeline. However, as discussed in \S\ref{sect:intro} the IMF of \piii stars is not yet constrained, and as discussed in \S\ref{meth:seds}, will determine whether \piii stars are detectable. 
\begin{figure*}
	\begin{center}
		\includegraphics[width=1.0\textwidth]{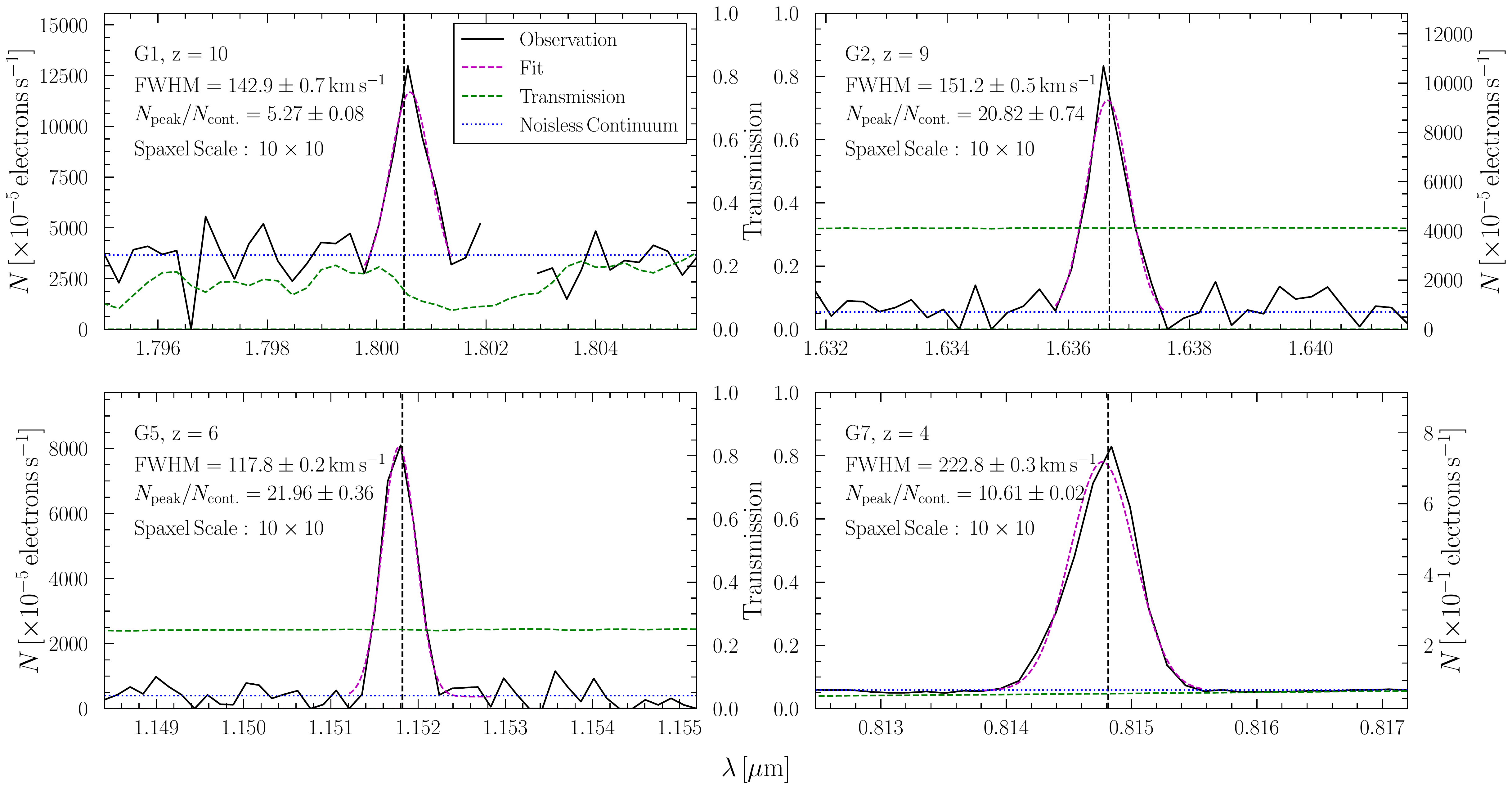}
		\caption{Same as Fig.~\ref{fig:lineprof} but for galaxies observed when using the $\rm PopIII.2$ IMF. Only G1, G2, G5 and G7  are shown as these are the only galaxies that produce an observable emission line with this IMF.
			}	
		\label{fig:m2}
	\end{center}
\end{figure*}
In the following sections we explore how two additional IMFs (${\rm PopIII.2}$ and ${\rm PopIII.K}$, see \S\ref{meth:seds} and Fig.~\ref{fig:sed}) impact the detection of our sample of simulated galaxies. We note that for the ${\rm PopIII.2}$ and ${\rm PopIII.K}$ IMFs we find a general trend for $L_{1640\lambda}$ to increase with Final $\sfr$ is found.

When changing from ${\rm PopIII.1}$ IMF to either the ${\rm PopIII.2}$ or ${\rm PopIII.K}$ IMF only the SED of every star particle is changed. All other properties of the gas and star particles of our sample of  galaxies remain constant. Therefore any change in line strength must be a result of changing the IMF.
\subsubsection{${\rm PopIII.2}$ IMF}
When employing the ${\rm PopIII.2}$ IMF all eight galaxies produce an intrinsic \heii emission line before observations, (values of $L_{1640\lambda}$ are given in column 3 of Table~\ref{table:lumi}). However these lines are significantly weaker, for example G2, when assuming the ${\rm PopIII.1}$ IMF has $L_{1640\lambda}\sim1.7\times10^{42}\,{\rm erg\,s^{-1}}$ but this falls to $\sim3.6\times10^{41}\,{\rm erg\,s^{-1}}$ when using ${\rm PopIII.2}$. In the cases of G6 and G3 this decrease in the intrinsic luminosity results in the galaxies no longer being detectable when observed with HARMONI (via \hsimp). 

Unlike the other six galaxies, G3 and G4 increase their intrinsic luminosity (i.e. $L_{1640\lambda}$) when using the ${\rm PopIII.2}$ IMF. This is a result of the ``dead'' star particle criteria when constructing the SSDC (see \S\ref{meth:stars}). For example, $\sim37\%$ of G4's stellar population have  $t_{\star,\rm age}>t_{\rm SN}$ when using ${\rm PopIII.1}$, however this falls to $\sim31\%$ when using ${\rm PopIII.2}$. These additional stars, though old are thus able to contribute additional ionising photons which increase the luminosity of the emission line. 

\begin{figure}
	\begin{center}
		\includegraphics[width=0.5\textwidth]{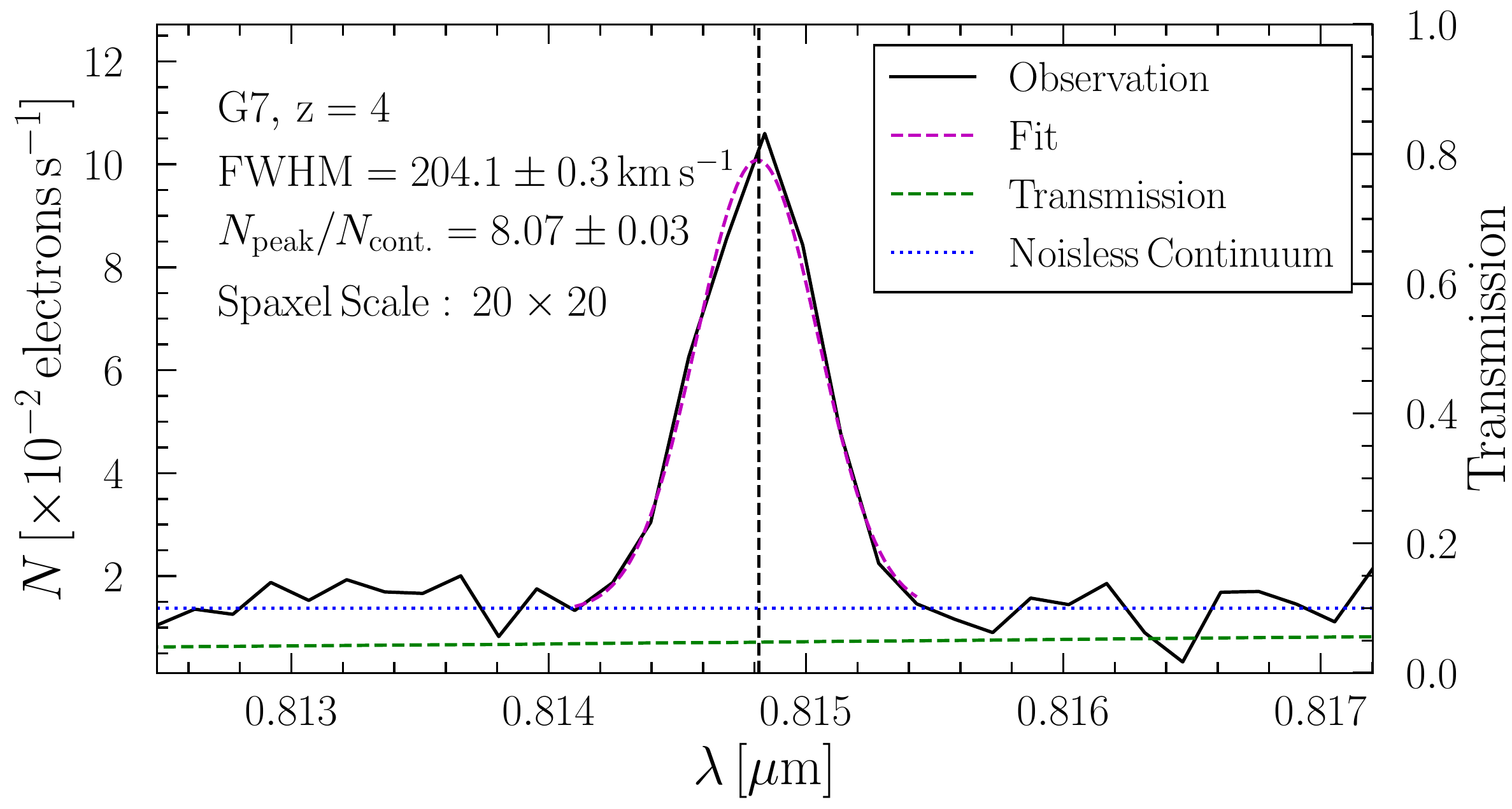}
		\caption{Same as Fig.~\ref{fig:lineprof} but for galaxies observed when using the $\rm PopIII.K$ IMF. Only G7 is shown as it is the only galaxies that produce an observable emission line with this IMF.
			}	
		\label{fig:mK}
	\end{center}
\end{figure}

Despite the ${\rm PopIII.2}$ IMF reducing their intrinsic luminosity, galaxies G1, G2, G5 and G7 still produce a sufficient number of photons in the \heii emission line that after observation with \hsim the line is detected (see Fig.~\ref{fig:m2}). In these four galaxies the post-observation line strength has (as expected) also decreased and in all but one case we find so too has  $N_{\rm peak}/N_{\rm cont.}$. In the case of G5, while the overall line strength has decreased, $N_{\rm peak}/N_{\rm cont.}$ has increased (see text on Fig.~\ref{fig:lineprof} and \ref{fig:m2}). To explain this increase we looked at both the noiseless output from \hsim and the pre-observation spectra of G5 for both IMFs and find that when switching from ${\rm PopIII.1}$ to  ${\rm PopIII.2}$ there is a decrease in the continuum strength. In the case of the former IMF the pre-observation continuum has magnitude of a few times that of the noise modelled by \hsimp, while in the latter case the continuum has a magnitude approximately equal to the added noise which results in a much weaker post-observation continuum and hence larger $N_{\rm peak}/N_{\rm cont.}$.

\subsubsection{${\rm PopIII.K}$ IMF}

Finally we consider the ${\rm PopIII.K}$ IMF (see Eq.~\ref{eq:imf2} and Fig.~\ref{fig:imf}). As before when switching from the ${\rm PopIII.1}$ to ${\rm PopIII.2}$, we find a reduction in the pre-observation line strength of all our galaxies when switching from ${\rm PopIII.2}$ to ${\rm PopIII.K}$, due to the lack of massive ($>100\Msol$) stars, for example, G2 when using this IMF has $L_{1640\lambda}\sim4.1\times10^{40}\,{\rm erg\,s^{-1}}$. The strength of emission lines in most galaxies is too weak to be detected after observations with \hsim  with only G7 still producing a detectable line, shown in Fig.~\ref{fig:mK}. Given that G7 is the brightest galaxy in our sample by more than an order of magnitude, independently of the choice of IMF and that when using the ${\rm PopIII.K}$ IMF it has a $L_{1640}$ which is $\sim3$ times brighter than G1 with the ${\rm PopIII.2}$ IMF (see Table~\ref{table:lumi}) it is no surprise that this galaxy would be detectable with any of the three IMFs.  However G7 appears to be a special case, as it not only has the largest stellar mass, average \sfr\, and final \sfr\, but it also appears to be the most chemically pristine galaxy (despite being at $z\sim4$) of our sample. 
 We therefore conclude that if \piii stars do follow the same IMF as \pii and \pl stars (i.e. a Kroupa-like IMF) it is highly unlikely that we will be able to observe and detect them via this emission line, unless galaxies with a stellar mass of a few $10^{7}\Msol$ and $Z\lesssim0.001Z_{\odot}$ are common at $z=4$.
 
It is worth noting that changing the observational setup, e.g. a longer exposure time or switching to the $20\,{\rm mas}$ pixels, does allow for the detection of the \heii emission line from other galaxies in our sample such as G6.

\subsection{Distinguishing Different IMFs}
\label{res:dimfs}

\subsubsection{Using \ha to Determine the IMF}
\begin{figure}
	\begin{center}
		\includegraphics[width=0.47\textwidth]{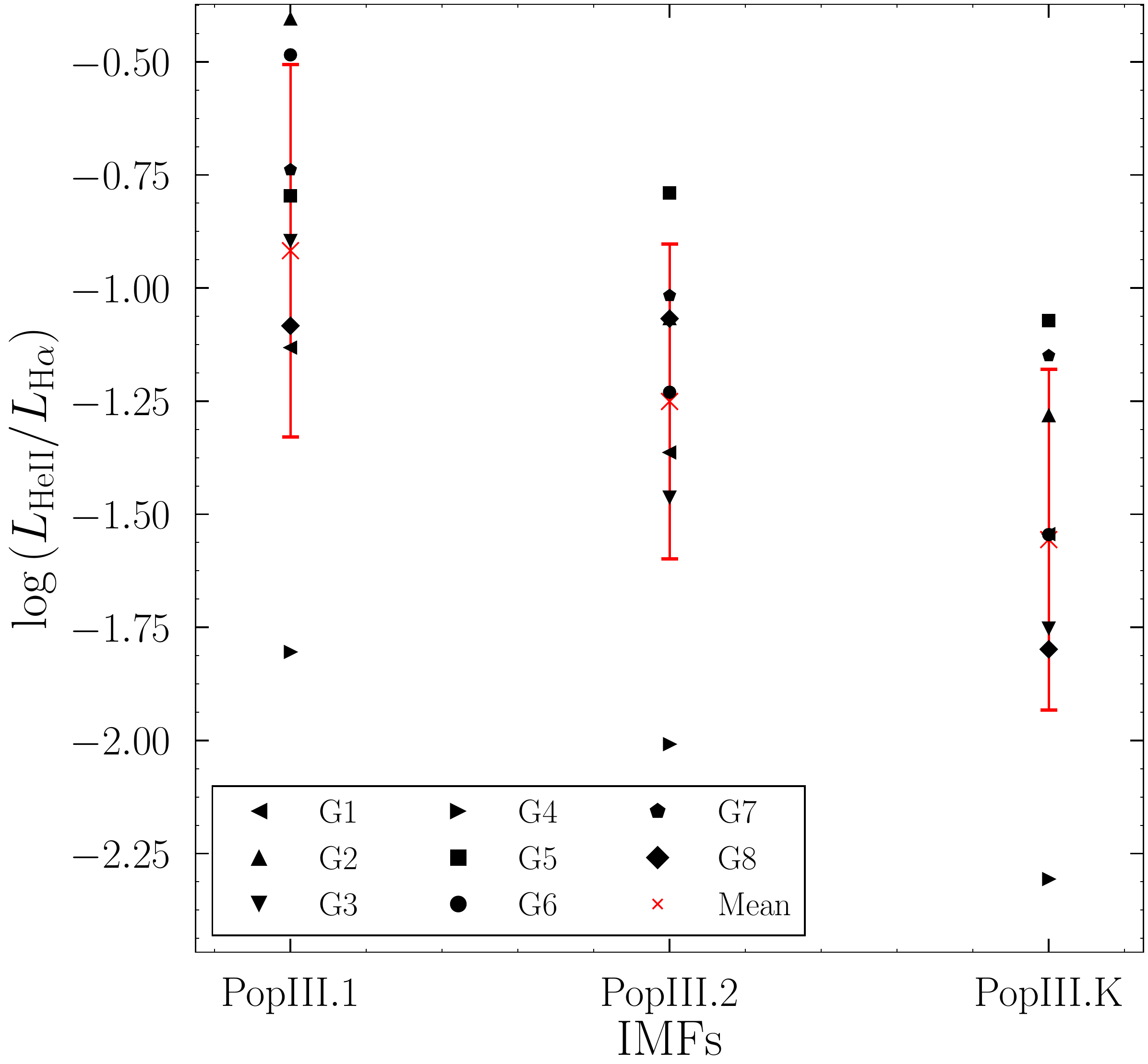}
		\caption{Ratio of the \heii and \ha luminosities for our eight sample galaxies. For each galaxy the ratio is shown when calculated for each of the three IMFs. The black points indicate the ratio of a given galaxy with a given IMF (see legend for details), while the red cross and error bars show the mean value and the standard deviation for each IMF.}		
		\label{fig:ratios1}
	\end{center}
\end{figure}

We now turn to the question of determining which of these IMFs is found in the Universe. One possible method for distinguishing between these three IMFs is comparing the relative strength of \heii and  \ha ($\lambda\sim6563\ang$) emission lines. Unfortunately at the redshifts of interest for detecting \piii stars the \ha line does not fall within the wavelength range of HARMONI. As a result any comparison between \ha and \heii would require a coordinated approach using multiple observing facilities, e.g. observing the same object with both HARMONI on the \eelt combined with observations with James Webb Space Telescope (JWST). 

\hsim is designed to mimic HARMONI and does not allow for mock-observations of emission lines outside of HARMONI's wavelength range. We do not therefore produce mock-observations of \Ha,  instead our discussion here focuses on the pre-observation SSDCs centred on \heii and \Ha.\footnote{Our pipeline is able to produce SSDC centred on any line available in \cloudy.}. We take the SSDCs for all eight galaxies with each IMF, calculate single aperture, continuum subtracted, \heii and \ha spectra. By integrating over these the total luminosity of the \heii ($L_{\rm He{\small II}}$) and \ha emission lines for each galaxy (with each IMF) is found. Fig.~\ref{fig:ratios1} shows how $\log\left(L_{\rm He{\small II}}/L_{\rm H\alpha}\right)$ varies for each galaxy as we change the IMF. 

In general $\log\left(L_{\rm He{\small II}}/L_{\rm H\alpha}\right)$ decreases as we move from ${\rm PopIII.1}$ to ${\rm PopIII.K}$. We note that the change in $\log\left(L_{\rm He{\small II}}/L_{\rm H\alpha}\right)$ when moving from one IMF to another is not identical for each galaxy, i.e. G2 has the largest value when using the ${\rm PopIII.1}$ IMF while G8 has its largest values when using the ${\rm PopIII.2}$ IMF. To get an estimate of how this ratio varies with IMF, the mean value and the standard deviation are also plotted, again finding a trend of a decreasing value as the IMF moves from ${\rm PopIII.1}$ to ${\rm PopIII.K}$, i.e. we calculate the means and standard deviations to be $-0.917\pm0.412,\,-1.251\pm0.348$ and $-1.556\pm0.377$. The mean value for a given IMF falls within the error bars of either of the other two, which is most likely the result of our limited data set (i.e. only 8 galaxies). It is possible as we expand the catalogue of mock observations that it will be easier to statistically separate the different IMFs using this ratio.  

\subsubsection{Using \lya to Determine the IMF}
\label{dis:lya}

\lya ($\lambda=1215.67\ang$) is often the brightest emission line from high-z galaxies, and it is likely that most targets observed with HARMONI will have been discovered through observations of their \lya emission, with follow up observations in \heii  \citep{Cooke:2009aa}.
As HARMONI can observe the \lya line for most \piii targets, it is interesting to simulate whether the flux of the \lya emission can provide additional information to constrain the Pop III IMF, for example can the ratio of \heii to \lya be used as an additional method for differentiating the IMF of \piii stars. 
One of the key advantages of \lya over \ha is that the former falls within HARMONI's wavelength range. This means for galaxies at some redshifts (e.g. $3, 6$ and $7$) observations of both lines can be carried out simultaneously as both lines fall within the same grating setting. Sadly this is not true for all wavelengths and multiple observations of a given target will still need to be carried out, but with the same instrument (HARMONI). 

Unfortunately, \lya is a resonant line, and \lya photons emitted from one part of the galaxy can be absorbed by gas in another part of the galaxy, and removed from our line-of-sight. This results in asymmetric line profiles, which are commonly observed \citep[e.g. see Fig.~3 of ][]{Sobral:2015aa}. Furthermore, the exact fraction of photons removed from our line-of-sight due to resonance effects is highly dependent on source geometry and kinematics, and cannot be quantified observationally which makes interpretation of observations at high redshift difficult \citep{Laursen:2007aa,Hoag:2019aa}.
Additional complications arise to \lya photons interacting with the IGM between the emitting galaxies and the observer. The reduction in \lya photons by the IGM scattering is dependent upon the reionisation history and the shape of the \lya emission line after resonant scattering with neutral hydrogen in the ISM and CGM. This can result in as little $10\%$ of the photons being transmitted through the IGM to observers \citep[][]{Dijkstra:2007ab,Zheng:2010aa,Dayal:2011aa,Laursen:2011aa}.  
For a more complete discussion on the difficulties arising from the resonant nature of \lya and how this is combatted in modern observations we refer the reader to \cite{Dijkstra:2014aa} and reference within.

\begin{figure}
	\begin{center}
		\includegraphics[width=0.47\textwidth]{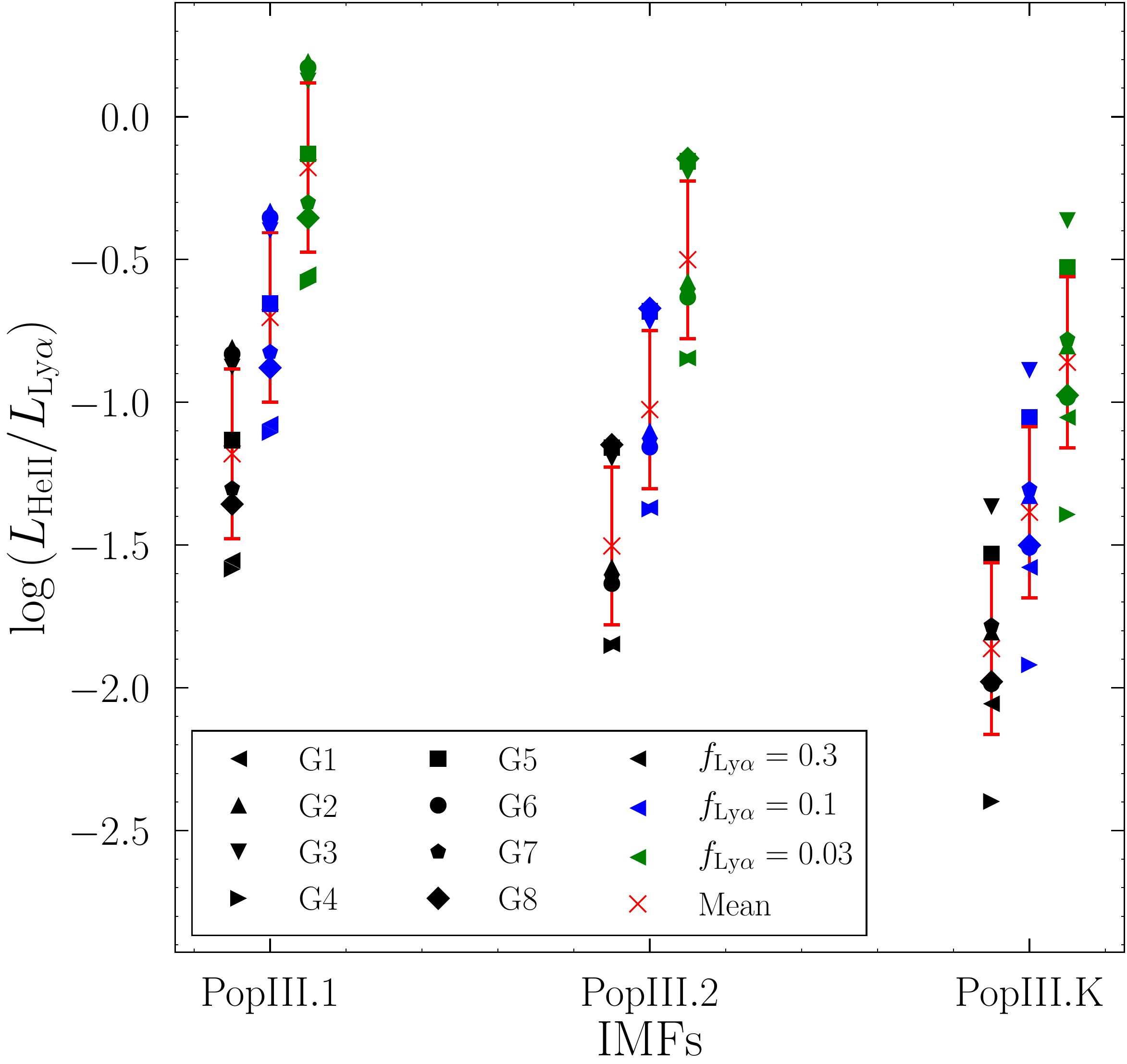}
		\caption{Same as Fig.~\ref{fig:ratios1}, but showing the ratio of \heii to \lya emissions. For each IMF the three different values of $\flya$ are given with the largest value on the left and the smallest on the right. These are also coloured to match the legend on the figure. 
		}		
		\label{fig:ratios2}
	\end{center}
\end{figure}

\begin{figure}
	\begin{center}
		\includegraphics[width=0.47\textwidth]{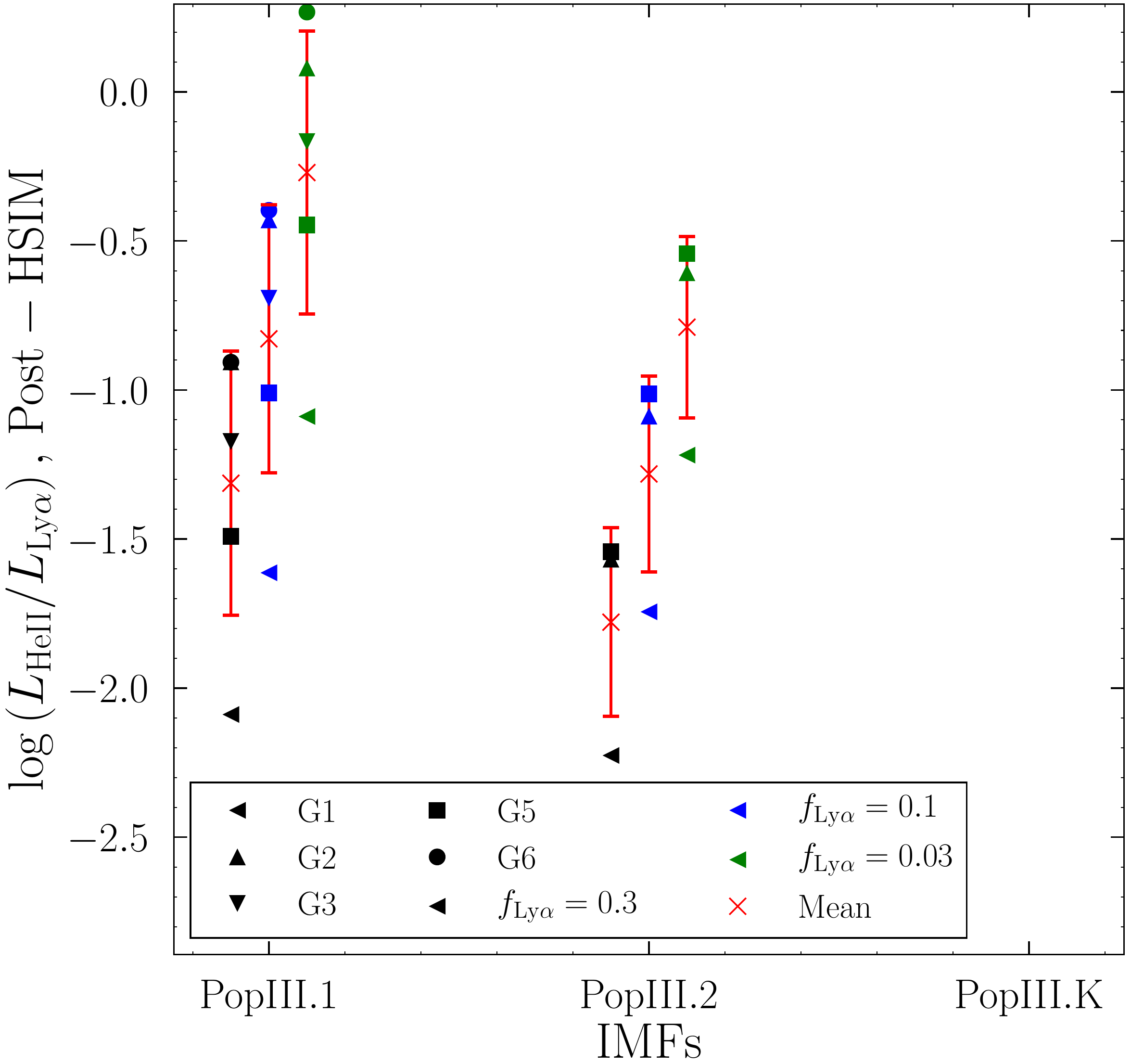}
		\caption{Same as Fig.~\ref{fig:ratios2}, but showing the ratio of \heii to \lya emissions after observation with \hsimp. Only galaxies that produces an observable \heii line are included. Due to the exclusion of G7 there are no detectable galaxies post observation for the ${\rm PopIII.K}$ IMF. 
		}		
		\label{fig:ratios3}
	\end{center}
\end{figure}

Modelling the above effects is beyond the scope of the current work. As the fraction of \lya photons removed from our line of sight varies with the content of the IGM, we use 3 different values of $\flya\,(0.03,\,0.1$ and $0.3)$ to quantify the impact $\flya$ for a range of possible sight lines. We define $\flya$ as the ratio of \lya photons escaping the galaxy along our line of sight to the number that would have been seen without any resonance effects, dust extinction, etc. The above values of $\flya$ are chosen to cover the range inferred from observations  \citep[][]{Hayes:2011aa,Blanc:2011aa,Dijkstra:2013aa} and simulations \citep[][]{Garel:2020aa}.

Our processing pipeline calculates the \lya emission following the methods described above for the \heii emission line (see \S\ref{meth:seds} \& \ref{meth:spec}), however before observing the \lya emission with \hsim we reduce its line strength by $\flya$ while keeping the continuum unchanged. 
While the \lya emission line falls within the wavelength range covered by HARMONI for objects at $3\leq z\leq10$, LTAO is not available to observe \lya for $z < 4.3$ (i.e. observed wavelengths shorter than $\sim6500\ang$). Therefore, we excluded G7 and G8 from our post-observation analysis of \Lya.
As the \lya emission line is not the primary focus of this work and due to the simple order of magnitude model used to account for resonant effects we do not present our recovered emission line profiles as results, however we have included them in Appendix \ref{app:lya} for the reader's convenience. 

For each galaxy, with each IMF and choice of $\flya$ we calculate $L_{\rm Ly\alpha}$ and thus $\log\left(L_{\rm He{\small II}}/L_{\rm Ly\alpha}\right)$ from the SSDCs, see Fig.~\ref{fig:ratios2}. For a given $\flya$ we find, as with the \ha ratio, that $\log\left(L_{\rm He{\small II}}/L_{\rm Ly\alpha}\right)$ tends to decrease when moving from a very top heavy IMF to a standard Kroupa IMF. This trend is also seen in the mean and standard deviation calculated for each data set. For a given IMF there is a clear offset between $\log\left(L_{\rm He{\small II}}/L_{\rm Ly\alpha}\right)$ calculated with $\flya=0.3$ and $0.03$, this is entirely a result of the choice of $\flya$. 

Fig.~\ref{fig:ratios3} shows the recovered values of $\log\left(L_{\rm He{\small II}}/L_{\rm Ly\alpha}\right)$ after the galaxies are observed with \hsimp, however only galaxies that produce a post-\hsim detectable \heii emission line are included in this figure. The units of the spectra are converted from electrons per second to ${\rm erg\,s^{-1}}$ before the ratio is calculated.  We again see the same trend of decreasing ratios as we move from ${\rm PopIII.1}$ to ${\rm PopIII.2}$ as well as the vertical displacement of points for a given IMF but different $\flya$. As before we confirm that this displacement is purely the result of $\flya$. 
The fact that our ``observations'' are of very faint objects means that we are measuring lines close to the noise limit and thus the ratio for a given galaxy can be different after observation. We find a typical error on our ratios of $\sim0.03$. This is highlighted by large scatter for galaxies using ${\rm PopIII.1}$ and $\flya=0.03$ compared to observations of the same objects with $\flya=0.3$, e.g. the \lya emission line is almost completely lost in noise for G6 when using ${\rm PopIII.1}$ and $\flya=0.03$ (see Fig.~\ref{fig:lyalinem1}) and we find $L_{\rm He{\small II}}> L_{\rm Ly\alpha}$.

As with \ha we are limited by the size of the data set when drawing definitive conclusions as to whether $\log\left(L_{\rm He{\small II}}/L_{\rm Ly\alpha}\right)$ can be used to determine the actual IMF of \piii stars. That being said our results suggest that such a ratio would be a promising test, with one caveat: $\flya$ must be known. For example if we compare the pre-\hsim ratios for ${\rm PopIII.1}$ and $\flya=0.3$ and ${\rm PopIII.K}$ and $\flya=0.03$ (which have mean values of $-1.180\pm0.297$ and $-0.860\pm0.299$ respectively) they are consistent with each other. Likewise, ``observation'' of G7 with ${\rm PopIII.K}$ and $\flya=0.03$ is consistent with the same galaxy using the  ${\rm PopIII.1}$ IMF and $\flya=0.3$ Therefore, in order for this ratio test to be used as a reliable method for determining the IMF of Pop III stars, further knowledge on $\flya$ would be required.

\section{Discussion}
\label{sect:disc}

\subsection{Other Source of the \heii emission Line}
\label{disc:agn}
In this work we have focused on determining whether it will be possible to use HARMONI on the \eelt to detect \piii stars using the \heii emission line, however these stars are not the only source of this line. As discussed in \S\ref{sect:intro} other luminous objects are also capable of producing the \heii emission line \citep[see also Table~8 of][]{AllensAstro}.  It is therefore important to have diagnostics of this emission line that provide a determination as to the type of object producing the emission line. 

 For example, the \heii emission line from WR stars should produce a broad emission line \citep[i.e. FWHMs $>800\kmsec$,][]{Leitherer:1995aa, Cassata:2013aa}. We predict that \piii stars will produce \heii emission lines with $80\leq {\rm FWHM}<225\kmsec$ (see Fig.~\ref{fig:lineprof}, \ref{fig:m2} and \ref{fig:mK}). Therefore measuring the line width of any observed \heii lines provides a simple way to distinguish between WR and \piii stars. 

Naively, one might expect that AGN emissions should predominantly impact the centre of a given galaxy \citep{Zeilik}. Therefore, a sufficiently bright extended source which is spatially resolved may allow us to distinguish between \piii stars and AGN. As shown by Fig.~\ref{fig:maps}, the current sample of simulated galaxies are not spatially resolved and cannot be used to distinguish between these two types of object. Recent observations \citep{Reines:2019aa} and the modelling of supermassive black holes in \newhorizon now indicate that AGN are not always centrally located. As a result spatial location and compactness alone is not enough to rule out the possibility of the signal being from an AGN and additional diagnostics are required. 

Due to their SED, the spectra of AGN are able to produce both the N{\sc v} doublet ($\lambda=1238.82$ and $1242.80\ang$) and O{\sc iii} ($1665.85\ang$) emission lines (see appendix \ref{app:agn}) if the accreting gas is sufficiently metal rich. These lines should not be present in spectra from \piii stars in part due to the shape of the SED and because of the low metallicity of the surrounding gas. Therefore the presence of these lines would be a strong indicator that the \heii emission line is due to an AGN and not \piii stars. In future work we will make use of the AGN modelled in the \newhorizon simulation to explore how the presence of AGN will impact the ability of HARMONI to detect \piii stars. 

\begin{figure}
	\begin{center}
		\includegraphics[width=0.48\textwidth]{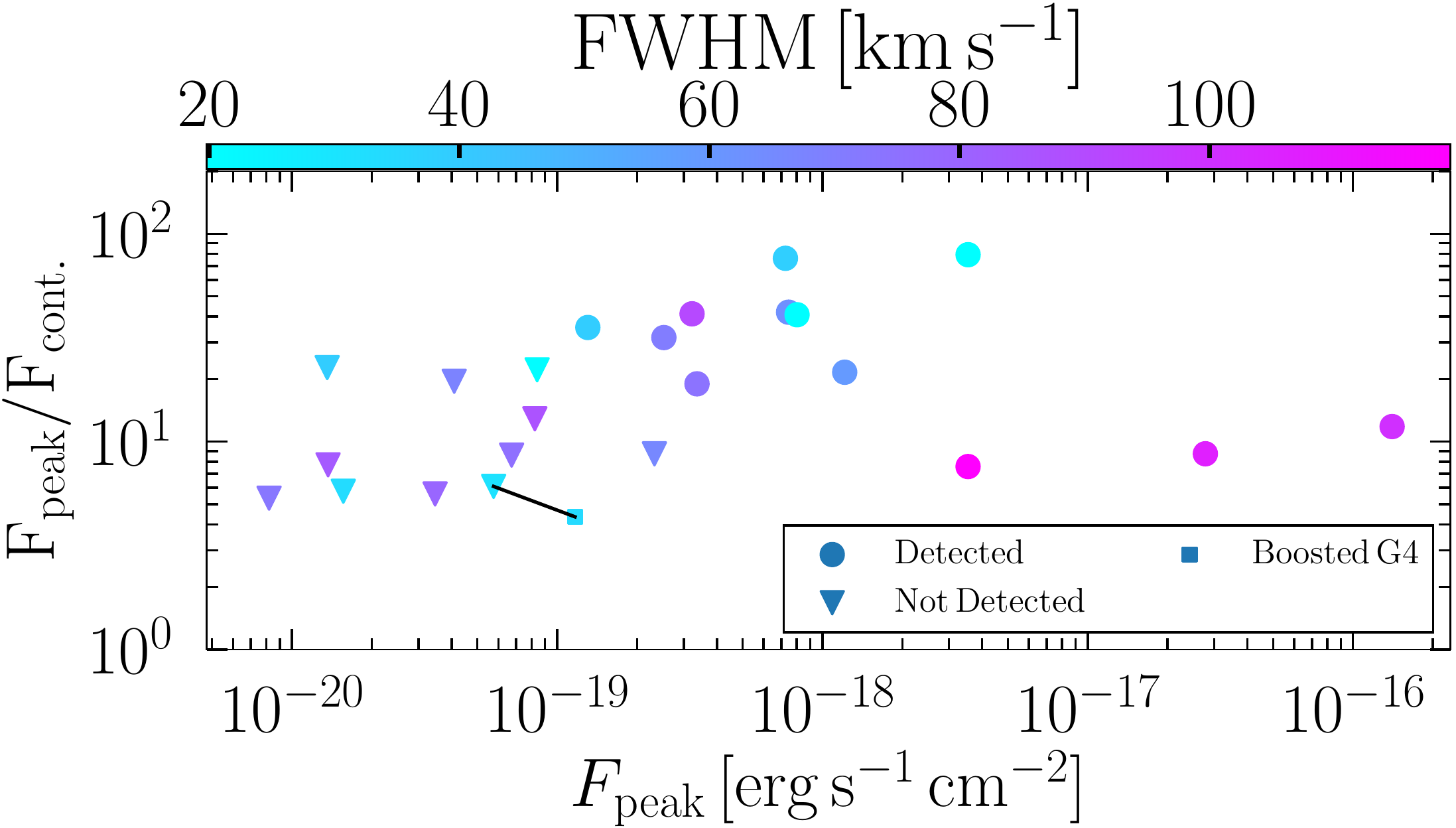}
		\caption{Comparison of pre-observation properties ($F_{\rm peak}$, $F_{\rm peak}/F_{\rm cont.}$ and line FWHM) of the \heii emission line for all galaxies (excluding G4 with the ${\rm PopIII.K}$ IMF). Galaxies with a detectable \heii emission line after observation with \hsim are shown by circles, while those that are not are marked by triangles. The single square point marks the values found for galaxy G4 after the brightness of each star particle is increased by a factor of ten. The black line links this 'brightened' G4 measurement to its original to aid in comparison.
		}		
		\label{fig:det}
	\end{center}
\end{figure}
\subsection{Detection Threshold}
\label{disc:det}

We now explore what determines the detectability of \piii stars in galaxies. The simplest answer is the total number of (young) \piii stars producing a large number of  photons able to doubly ionise He and thus produce a strong \heii emission line, but how does this manifest in observations? 
By assuming an IMF that is even more top heavy than {\rm PopIII.1} we are are able to increase the luminosity of each star particle by a factor of ten. This in turn increases the magnitude of the SED for each particle. Alternatively, \cite{Mas-Ribas:2016aa} found that if stochastic sampling of the IMF is properly accounted for and with the correct physical conditions, emission lines form \piii stars can be up to an order of magnitude brighter. 

To explore the result of increasing the luminosity of star particles it is useful to define two terms: $F_{\rm peak}$ as the maximum value of the \heii emission line (before observation with \hsimp) in the a single aperture spectrum similar to those shown in Fig.~\ref{fig:lineprof} and $F_{\rm cont.}$ the estimated value of the continuum at same wavelength as $F_{\rm peak}$. For both quantities we take their values in units of spectral brightness (i.e. $\lunittl$) and multiply this value by the spatial and spectral dimensions of a spaxel (e.g. $10\mas\times10\mas\times0.5\ang$), thus both $F_{\rm peak}$ and $F_{\rm cont.}$ are fluxes in units of $\lunit$. As both are in the same unit we are able to calculate the unit-less ratio of those two values: $F_{\rm peak}/F_{\rm cont.}$, which we use as one of our diagnostic tools. 

Increasing the luminosity of each star particle, as expected, results in a brighter galaxy.  For example, the change in the brightness of a single star particle's SED by a factor of ten, increase their output from CLOUDY and thus the galaxy's single aperture line brightness is larger (i.e. $F_{\rm peak}$ increases), typically by a factor of a few. The exact increase in brightness depends on the star particles host cell's gas density and can results in small variations in  $F_{\rm peak}/F_{\rm cont.}$. For G4 increasing its brightness by a factor of ten is \emph{sufficient} to make the galaxy observable with the ${\rm PopIII.1}$ IMF, in this case $F_{\rm peak}$ is increased from $5.8\times10^{-20}$ to $1.17\time10^{}{-19}\lunit$.

Fig.~\ref{fig:det} further explores what determines the detectability of the \piii stars via the \heii emission line by comparing the (pre-observation values) of $F_{\rm peak}$, the emission line's FWHM and $F_{\rm peak}/F_{\rm cont.}$ for all the galaxies. We find that $F_{\rm peak}$ appears to be the most important factor in ensuring a line is detected. Indeed, both the ratio of $F_{\rm peak}/F_{\rm cont.}$ and the FWHM have little impact on the detectability of \Heii: i.e. both detected galaxies and undetected galaxies have $5\lesssim F_{\rm peak}/F_{\rm cont.}\lesssim80$ and $19\lesssim $FWHM$\lesssim120$. G9, when using the ${\rm PopIII.2}$ IMF, has the lowest value of $F_{\rm peak}$ for a detected galaxy thus we use this galaxy to set our lower limit for detection, i.e. $F_{\rm peak}>10^{-19}\lunit$. Furthermore, during the brightness test above, G4 became detectable once $F_{\rm peak}>10^{-19}\lunit$ despite a drop in $F_{\rm peak}/F_{\rm cont.}$, thus providing further support for $F_{\rm peak}>10^{-19}\lunit$ being the detection threshold.

To remain self gravitating smaller galaxies must have small velocity dispersion (on the order of $100\kmsec$). This leads to emission lines with smaller FWHM and thus the  photons from the line are more concentrated, boosting $F_{\rm peak}/F_{\rm cont.}$ and leading to easier detections. So while the range of FWHM found for the high redshift galaxies in our sample does not seem to impact detectability, it is likely FWHM would play a role for galaxies with larger velocity dispersions.

Finally, when selecting galaxies from \newhorizonp, we specifically selected compact objects with a high percentage of \piii stars (see \S\ref{meth:nh:sel}) as this presented the best chance for successful observations. Given that (by design) the simulation produces a volume that matches the average density of the Universe and as a result, at $z=0$ the most massive halo produced in \newhorizon will have a mass on the order of $\sim10^{13}\Msol$. Therefore the galaxies formed in the simulation \emph{should} represent the galaxies most likely to be in the Universe. Rarer and larger objects such as the Virgo Cluster, with mass on the order of $10^{15}\Msol$ \citep{Fouque:2001aa}, will not be formed.  From this, we infer that for a given sight line it is likely that at least one \heii bright, chemically pristine galaxy at $3\leq z\leq10$ that meets the detection limits described above and our selection criteria will be found.

%
\subsection{Validity of Approach}
\label{dis:valid}

\subsubsection{Post Processing vs. On The Fly}
The primary goal of this work is to determine whether \piii stars will be detectable with HARMONI and the \Eelt. To achieve this it has been necessary to adopt several assumptions. For example, we employed a metallicity cut to determine (in post processing) whether or not a star particle is a \piii star. However, normally a proper primordial chemistry network is used to model \piii star formation \citep[for example, see][]{Glover:2007aa,Jappsen:2009aa}.  If the goal of this work had been to model the formation of \piii stars and determine how they shaped the evolution of their host galaxy this would present a significant issue.

Furthermore such a study would require, in addition to the primordial chemical networks, that the feedback model employed in the simulation accounted for the top-heavy nature of the theatrical \piii IMFs. We emphasis this is not our goal, and as a result, we are able to make use the pre-calculated SEDs from {\sc yggdrasil} to account for the presence (or lack of) metals in \piii stars. The impact of metals in the ISM surrounding these stars is accounted for with the, well tested, radiative transfer program \cloudyp. This approach affords us the flexibility of being able to change the SED of a \piii stars at the time of analysis, allowing for the testing of multiple IMFs, in a variety of environments (i.e. gas densities, metallicities, etc) without having to rerun an extremely computationally expensive simulation. We therefore argue, the post-processing method used here is sufficient to provide a determination of \piii detectability with the \Eelt. 

\subsubsection{Large \piii Stellar Clusters}
 This work assumed that \piii stars form in clusters with a total mass $\geq10^{4}\Msol$. However, it is perhaps more likely that a single massive \piii star forms in a region and shuts down the local star formation via feedback. When this first star eventually undergoes a supernova it pollutes the local environment with metals and therefore the subsequent generation of stars will be too metal rich to be considered \piii \citep{Ostriker:1996aa,Omukai:1999aa}. In this scenario \piii stars are likely to be scattered throughout the galaxy and as a result fail to produce a strong enough emission line to be detected even with HARMONI and the ELT.

That being said, current work on the formation of $10^{5}\Msol$ black holes in the high-$z$ Universe provides a possible formation mechanism for clusters of \piii stars, as assumed in this work. One of the current methods for forming such black holes is ``direct collapse'' were gas clouds with masses of $\sim10^{5}\Msol$ collapse under gravity directly to a massive black hole \citep[see][and references within]{Begelman:2006aa}. These models require the gas to be pristine and unable to form H$_{2}$ which prevents the gas from cooling to below $10^{4}{\rm\,K}$. The lack of cooling stops the fragmentation into smaller clouds and instead a single massive black hole forms.  We assert that by relaxing the no-H$_{2}$ criteria the same arguments can be used to produce a cluster of \piii stars: the presence of H$_{2}$ results in the collapsing pristine cloud fragmenting and forming multiple \piii stars \citep[see][]{Abel:2000aa}. This model requires that the star formation happens throughout the cloud on time scales on the order of the free-fall time, as feedback (supernovae and ultraviolet radiation) from just one massive \piii star could disperse the cloud or dissociate the H$_{2}$. In essence, for a \piii cluster to be created, the star formation efficiency per free-fall time ($\epsilon_{\rm ff}$) must be on the order of unity. At $z=0$, giant molecular clouds shows $\epsilon_{\rm ff}$ ranging from $\sim10^{-4}$ to $\sim2$ \citep[see][and references within]{Grisdale:2019aa} and therefore finding clouds at high-$z$ with  $\epsilon_{\rm ff}\sim1$ is plausible. Even if not the norm, galaxies where this type of \piii formation occurs are the ones most likely to be detected with $30+{\rm\,m}$ class telescopes and therefore it is this kind of \piii star formation we focus in this work. 

Using radiation hydrodynamic simulations \cite{Susa:2006aa} explored the impact of one generation of \piii stars on subsequent generations. Finding that if a first generation \piii star has a mass of $120\Msol$, its ionising radiation creates a H$_{2}$ shell. This shields H$_{2}$ outside of the shell from the dissociating ultraviolet (UV) radiation produced by the star. As a result the star formation of the second generation stars will not be effected by the first. \cite{Hasegawa:2009aa} took this concept further, testing a range of different masses for the first generation \piii stars and found that star formation was only delayed if the first generation had a mass of $\lesssim25\Msol$, i.e. the ionising radiation was not sufficient to create a H$_{2}$ shell. Given that both the ${\rm PopIII.1}$ and ${\rm PopIII.2}$ IMFs have a mean/characteristic stellar mass of $\gtrsim100\Msol$, we argue that the star particles in this work could indeed form in this self-shielding regime and thus be a valid way of modelling \piii stars in cosmological simulation.

\cite{Xu:2016aa} showed that in the Renaissance Simulations not only did \piii stars formation occur down to $z=7.6$ but also that galaxies could continue to form multiple generations of \piii stars. This occurs as a result of two factors: (i) the local and slow nature of metal enrichment and (ii) the delay of \piii star formation due to Lyman-Wemer radiation from surrounding metal-enriched star formation. Given all of the above, we argue that our assumption of \piii star clusters is not as reasonable as it might initially seem.  

In this work we explored the detectability of \piii stars with HARMONI and the \eelt assuming that they follow one of three specific IMFs (see \S\ref{meth:seds} and Fig.~\ref{fig:imf}). While we have attempted to choose three IMFs that bracket the current theories on \piii IMFs, it is possible that \piii stars actually follow very different IMFs. For example, all IMFs in this work assume a continuously sampled mass function however the IMF might be instead be sampled stochastically \citep[see][for details]{Mas-Ribas:2016aa}. To properly determine the IMF of \piii stars, all possible, or at least a significant range of,  IMFs need to be considered and a catalogue of mock observations using methods such as those used in this work needs to be built up for comparison with observations.

\subsubsection{Are Hydro and \cloudy Simulations Really Needed?}

It could be argued that the methods used here are surplus to requirement and that SED modelling could produce the same results. However, this is only true to a limited extent and only because our chosen galaxies produce point-like sources when observed (see Fig.~\ref{fig:maps}). Indeed as mentioned above (see \S\ref{res:fmaps}) if we were to choose a more extended object, or one with a component of non-\piii stars, gas and stellar structures would become visible in integrated line strength maps both before and after observations. 
G6  demonstrates this (see Fig.~\ref{fig:maps}): the region we extract of the simulation contains two galaxies connected by a reasonably dense $(\sim100\Msol\pc^{-2})$ gas bridge. However due to the lack of stars in this bridge it is not visible in \heii either before or after observations with \hsimp. Furthermore, the \heii integrated line strength map (i.e. middle panel of Fig.~\ref{fig:maps}) shows that both of these galaxies have their \heii emissions spread out over more than $20$ pixels, with the in-falling galaxy (top galaxy in the panel) having elongated emission region which stretches over $\sim0.5\kpc$ yet once observed with \hsim both galaxy are reduced to just a handle of pixels. Indeed the previously elongated structure of the in-falling galaxy is completely lost (see the right panel for G6 on Fig.~\ref{fig:maps}). An identical behaviour is seen in G7. The lack of extended objects can partly be explained by the AO technique employed by the HARMONI, as it provides improved contrast on compact objects but provides little to no improvement to the detection of extend structures, such as those found in G6 and G7 \citep[][]{Davies:2012aa}. The AO, coupled with the decrease in like brightness in the extended regions (a result of the compact stellar structure of these early galaxies) provides an explanation as to why the extend regions and thus extended sources of \heii are unlikely to be detected.

The reader will remember that the spectrum recovered from the individual SSDCs for an individual galaxy are the result of hundreds or sometimes thousands\footnote{in our sample of eight galaxies the smallest has $\sim60$ star particles while our most massive has $\sim6000$!} of individual sources. The spectrum of each of these sources is determined by a number of factors including local gas density and the age of the star particle supplying photons, in short they are unique and combine to give each galaxy its own unique spectrum. Furthermore, the spectrum of each galaxy is shaped by its evolutionary history in the simulated universe of \newhorizonp,  which was designed to be a realistic model of our Universe \citep[see][]{Park:2019aa,Dubois:2020aa}. Therefore the SSDCs and the spectrum produced after observation with \hsim are cosmologically constrained for a given IMF. 

We argue the methods used in this work are the most realistic (i.e. cosmologically constrained) for the computational cost (i.e. without running full radiation-hydrodynamic simulation) and thus provide the closest mock observations of \piii stars in high $z$ galaxies to real observations. Finally, the post processing pipeline used here can be applied to any number of simulated galaxies, with any choice of IMF to build up a catalogue of mock observations for comparison with real observations made with HARMONI at the \Eelt.

While the work presented here is focused on the detectability of \heii emission lines from \piii stars, the pipeline developed for this work can be used in a variety of ways to prepare for $30+$ metre class telescopes, such as the \Eelt. For example, it would be possible to use cosmological simulations to identity regions at each $z$ within the simulated volume that are capable of forming \piii stars and galaxies. These regions could be processed with our pipeline and mock observations could be carried out. Such work might allow us to determine how proximity to non-\piii galaxies effects the line flux of  \piii galaxies.

\section{Conclusion}
\label{sect:con}

The existence of Population III (\Piii) stars has not been observationally confirmed up to now, although several attempts have been made, and some excellent candidates have been identified.  Given their primordial composition with no heavy elements, \piii stars are expected to be substantially more massive, as well as have poor coupling between their constituent gas and the emitted radiation.  Consequently, they should be much hotter, and have a much higher UV flux, capable of ionising not only Hydrogen but also Helium in the surrounding gas (H II region).  The strength of the \heii is thus a good observational diagnostic for the presence of Pop III stars, although the emission line will be faint due to the large luminosity distance of these very high redshift star forming regions. 
Starting with detailed cosmological simulations from the \newhorizon suite at a range of redshifts, and post-processing them with full radiative transfer and typical top heavy IMFs expected for Pop III stars, we have created self-consistent mock galaxies whose morphology, kinematics, line strengths and line-to-continuum ratios are representative of what \piii star dominated galaxies would exhibit at a range of redshifts.  Then, using HSIM, we simulated long (10 hour) observations of these mock galaxies with the world's largest telescope, the \Eelt, and its AO-assisted, first-light, integral field spectrograph, HARMONI.  The \Eelt's huge collecting area, coupled with the exquisite spatial resolution provided by HARMONI LTAO allows us to predict that the He II feature can be detected with good signal-to-noise from a substantial fraction of the mock galaxies, spanning a wide range of redshifts.  However, to be certain that the line indicates the presence of Pop III stars would require ancillary observations of the \ha line from these objects to measure the \he to \ha ratio, probably using the JWST, given the high redshifts involved.
Our key results are:
\begin{enumerate}
	\item If \piii stars have a top heavy IMF, with a mean mass $\geq100\Msol$ and a total mass of $\geq10^{4}\Msol$ they will produce strong \heii emission. In the case of galaxies with a compact \piii stellar distribution this line will be observable with HARMONI at redshifts ranging from ten to three. 
	
	\item While most of our sample galaxies are in essence point sources some do have structure prior to observation  with \hsim (e.g. G6 and G7). We find that the morphology of our galaxies is not recoverable from observations of the \heii emission line. We therefore advise that observations looking for \piii stars with HARMONI look for point-line sources rather than extended objects.
	
	\item If \piii stars follow an IMF similar to one found in current stellar populations, e.g. a Kroupa IMF, the galaxy will still produce a \heii emission line. However unless the emitting galaxy has a stellar mass of the order $10^{7}\Msol$, a very young stellar age and an average metallicity of $\lesssim0.001Z_{\odot}$ this line will be too weak to be observed. 
	
	\item \heii emission lines is observable in our simulations over a wide range of redshifts, only if line's peak flux is $>10^{-19}{\rm erg\,s^{-1}\,cm^{-2}}$ and if the emitting region is compact, i.e. a few tens of milliarcseconds across. 
	
	\item Determining which \emph{theoretical} IMF is close to the \emph{actual} IMF is very likely  possible by comparing the relative strength of other emission features, such as \ha and \lya. However, this will involve combining observation programs and possibly different facilities, such as the \eelt (for \lya and \heii emissions) and JWST (for \ha emissions). Furthermore, a sufficiently large data set will be required to allow for statistical analysis of the line ratios.
		
\end{enumerate}

In future work we will explore how signals from other objects capable of producing the \heii emission line will impact detections of the \piii stars and how these signals can be disentangled. 
 
\section*{acknowledgments}
We thank the referee for the constructive comments. 
KG and NT acknowledge support from the Science and Technology Facilities Council (grant ST/N002717/1), as part of the UK E-ELT Programme at the University of Oxford.
KG, JD and AS acknowledge support from STFC through grant ST/S000488/1.
The research of JD and AS is supported by Adrian Beecroft and STFC.
MPS acknowledges support from the Comunidad de Madrid through Atracci\'{o}n de Talento Investigador Grant 2018-T1/TIC-11035 and STFC through grants ST/N000919/1 and ST/N002717/1. 
TK was supported in part by the Yonsei University Future-leading Research Initiative (RMS2-2019-22-0216) and in part by the National Research Foundation of Korea (No. 2017R1A5A1070354 and No. 2018R1C1B5036146).
SKY acknowledges support from the Korean National Research Foundation (NRF-2017R1A2A05001116).\\
We thank all members of the New Horizon collaboration, and more specifically Sugata Kaviraj and Marta Volonteri for valuable comments and suggestions. \\
We thank Erik Zackrisson for fruitful and instructive guidance on the use of the Yggdrasil models.
Calculations were performed with version 17.01 of \cloudy \citep{Ferland:2017aa} and \hsim version 210 \citep{Pereira-Santaella:2019aa}

\section*{Data Availability}
The data underlying this article will be shared on reasonable request to the corresponding author.

\bibliographystyle{mn3e}
\bibliography{ref}

\appendix

\section{Analytical Model}
\label{app:ana}
Throughout this work we employ \cloudy to carry out all radiative transfer calculations, however as we are using non-standard SEDs it is important to determine if the results from \cloudy are reasonable. To this end, we run three \cloudy simulations (one for each \piii SED) for a $10^{4}$ year old star cluster with a total luminosity of $L_{\rm tot}=10^{38}{\rm erg\,s^{-1}},$\footnote{This $L_{\rm tot}$ corresponds to star cluster masses of  $1.7\times10^{2},\,5.0\times10^{2}$ and $3.0\times10^{3}\Msol$ for PopIII.1, PopIII.2 and PopIII.K SEDs respectively.} 
embedded in a cloud with a uniform density of H and He ($n_{\rm H}=100\cc$ and $n_{\rm He}=10\cc$).  To compare the results of the \cloudy tests we construct a simple analytical model to determine the expected strength of the \heii line. In the following we outline this model and compare it to the results from \cloudy.

\subsection{The Model}
\label{res:amod}
\begin{figure}
	\begin{center}
		\includegraphics[width=0.46\textwidth]{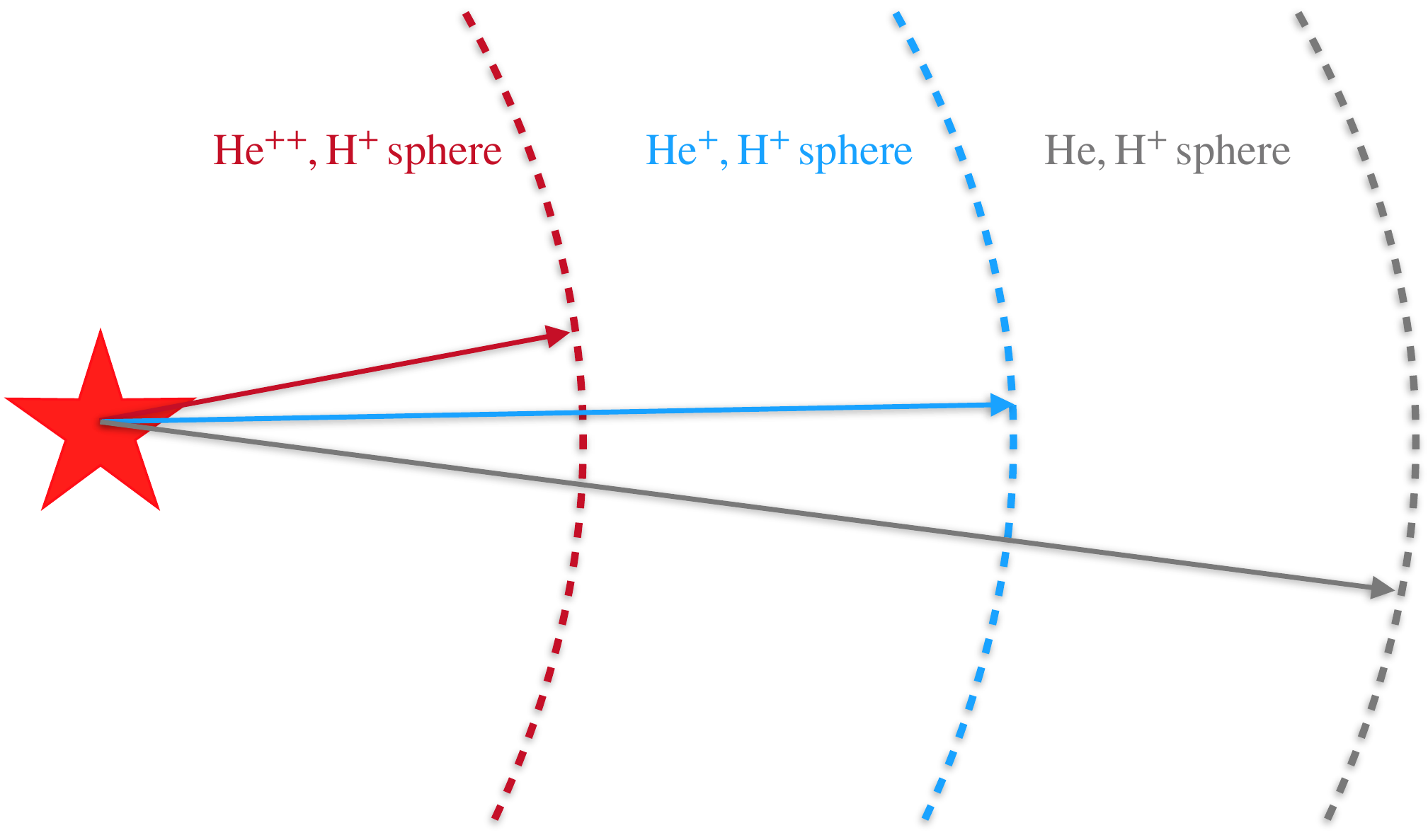}
		\caption{Cartoon depicting the analytical model, described in \S\ref{res:amod}. The star represents the stellar cluster, while the dashed lines show the boundaries, between the three regions (i.e. He$^{++}$ \& H$^{+}$, He$^{+}$ \& H$^{+}$ and He \& H$^{+}$) and the correspondingly coloured  arrows represent the radius of each Str\"omgren sphere.
		}
		\label{fig:amod}
	\end{center}
\end{figure}

Similar to \cloudyp, we treat the stellar cluster as a single point-like source of radiation embedded within a cloud with three distinct regions: a He$^{++}$ and H$^{+}$ inner sphere,  a surrounding shell consisting of He$^{+}$ and H$^{+}$ shell and an outer shell of He and H$^{+}$, as shown in Fig.~\ref{fig:amod}. The distance from the star cluster to the outer boundary of each of these regions can be estimated as a Str\"omgren sphere \citep{Stromgren:1939aa}, which is given by:
\begin{equation}
	R_{{\rm SS},i} = \left(\frac{3Q_{i}}{4\pi n_{i} n_{\rm e} \alpha_{{\rm B},i}}\right)^{1/3},
	\label{eq:ssphere}
\end{equation}
where $Q_{\rm i}$ is the number of photons per second with sufficient energy to ionise the chemical species $i$, $n_{i}$ is the number density of the species, $\alpha_{{\rm B},i}$ is the recombination rate of the species and $n_{\rm e}$ is the number density of electrons. 

The \heii emission can only be produced (in our model) in the regions where He is ionised i.e. $R\leq R_{\rm SS,He}$. We divide this region into concentric shells (of thickness ${\rm d}R$) and calculate the flux emitted by each shell using:
\begin{equation}
	j_{\lambda 1640} = \gamma j_{\lambda 4686} n_{e}  n_{\rm He}  {\rm d}R,
	\label{eq:linestrength}
\end{equation} 
where  $j_{\lambda 4686}$ is the strength of the He line at $4686\ang$ and $\gamma$ is the ratio of $j_{\lambda 1640}$ to $j_{\lambda 4686}$. Values for $j_{\lambda 4686}$  and $\gamma$ are taken from column 4 of Table 4.5 in \cite{Osterbrock:2006aa}. We adopt the simplification that the only ion that contributes to recombination is that of the species being considered, however the total number of available electrons needs to be considered therefore 
\begin{equation}
	n_{\rm e} = \left\{
	\begin{array}{l l} 
		n_{\rm H} &\quad\mbox{if $R_{\rm SS,He}<R\leq R_{\rm SS,H}$},\\
		n_{\rm H}+n_{\rm He} &\quad\mbox{if $R_{\rm SS,He^{+}}<R\leq R_{\rm SS,He}$},\\
		n_{\rm H}+2n_{\rm He} &\quad\mbox{if $R\leq R_{\rm SS,He^{+}}$}.\\
	\end{array}\right.
	\label{eq:ne}
\end{equation}

Finally, the total luminosity emitted at a given distance from the star cluster,  $L(\leq R)$, is calculated by summing the value of $j_{\lambda 1640}$ for each shell contained within $R$.

\subsection{Model Comparison}
\label{res:acom}

\begin{figure}
	\begin{center}
		\includegraphics[width=0.48\textwidth]{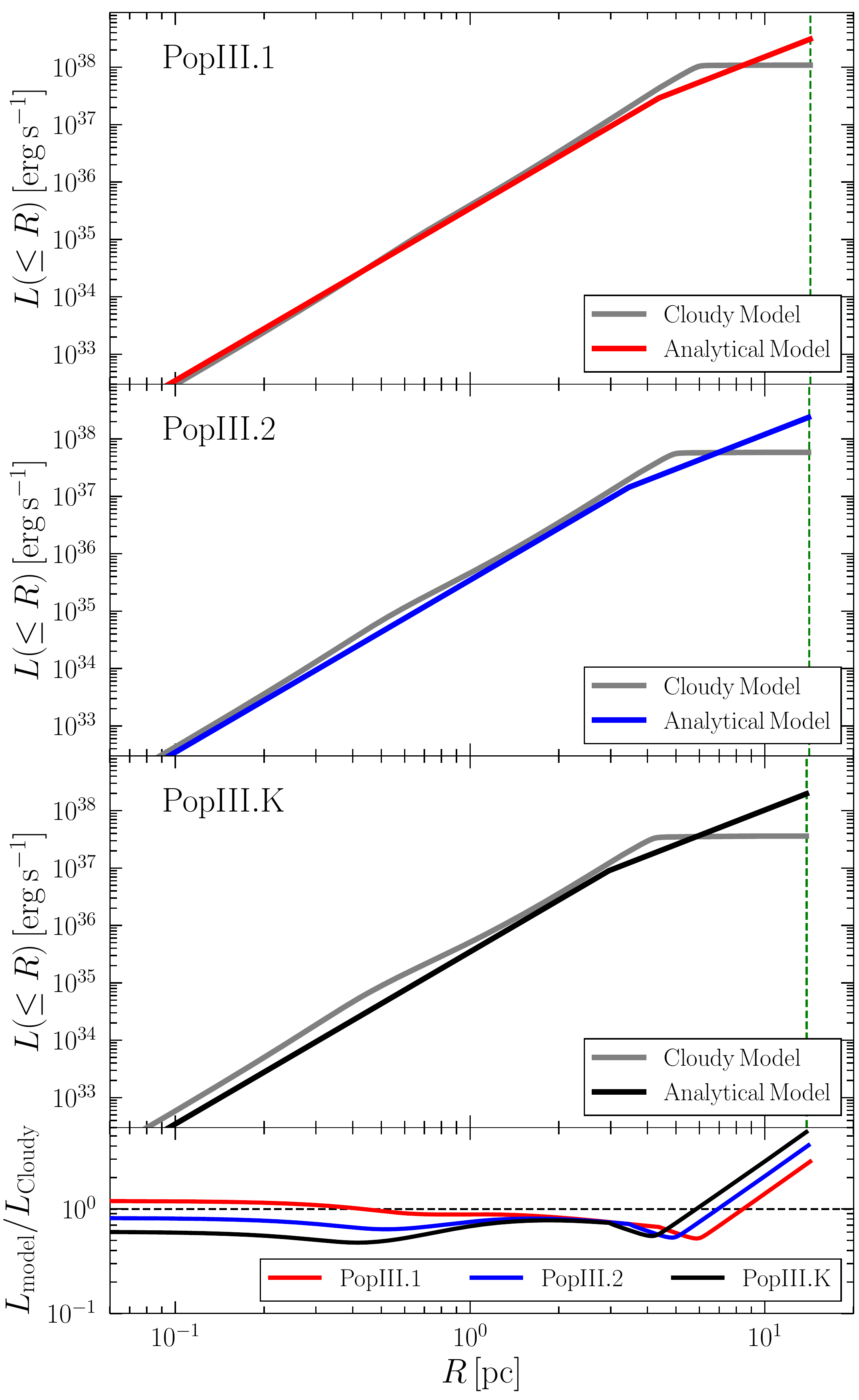}
		\caption{Comparison of the luminosity ($L(\leq R)$) at a given radius ($R$) from a star cluster as predicted by the analytical model and \cloudy simulations described in  \S\ref{res:amod}.  Shown are results for the PopIII.1, PopIII.2 and PopIII.K SEDs (red, blue and Black lines in top 3 panels), with the \cloudy values show in grey. The vertical dashed-green lines shows the He Str\"omgren radius calculated by the analytical method. The bottom panel shows the ratio of the analytical model and \cloudy values for $L(\leq R)$. The horizontal dashed-black line shows $L_{\rm model}=L_{\rm cloudy}$ 
		}
		\label{fig:amodr}
	\end{center}
\end{figure}

A comparison of the $L(\leq R)$ as calculated by the above model and \cloudy is shown in Fig.~\ref{fig:amodr} . We find that for most values of $R$ the two methods agree with each other to within a factor $\sim2$. It is worth noting that for $R\lesssim0.5\pc$ the analytical model tends to have a lower $L$ than \cloudy, while the opposite is true at larger $R$, as summarised in the bottom panel of Fig.~\ref{fig:amodr}. 

There are several differences between how \cloudy and the model outlined above compute $L(\leq R)$. The flattening seen in the \cloudy model at $R\sim4-5\pc$ corresponds to the radius as which only $\sim 40-80\%$ of He$^{+}$ is ionised, i.e. Str\"omgren sphere calculated by \cloudy. The Str\"omgren sphere as calculated from the analytical model (shown in Fig.~\ref{fig:amodr} by the green dashed line) is larger than the sphere calculated by \cloudy. This difference in size of Str\"omgren sphere arises from \cloudy taking into account cooling processes that are neglected in the analytical model as a result the latter assumes a constant temperature throughout the gas. Furthermore the analytical model assumes that hydrogen atoms have no impact on the number of photons available to create He$^{+}$ and similarly for ions He$^{+}$ have no impact on the number He$^{+}$ ionising photons. In reality this is not the case and more complete treatment than that carried out in our model is need to completely account for the impact of different species. 

Due to the overall agreement between our analytical model and \cloudy in the above test case, we are confident that \cloudy can be used as outlined in \S\ref{meth:spec} to generate spectra for our simulated galaxies.

\section{\lya Emission Line Profiles}
\label{app:lya}

For the reader's convenience we provide the post-\hsim \lya spectra of our simulated galaxies here. How these lines are produced and the significance of $\flya$ is outline in \S\ref{dis:lya}. In this work we only compare the \lya emission lines with $\flya=0.03,\,0.1$ and $0.3$ to the  \heii emission line, but here we include the emission line with $\flya=1.0$. In the following figures we only show galaxies at $z>4.3$ (i.e. G7 and G8 are excluded) and for which we can detect a corresponding \heii emission line. 
\begin{figure*}
	\begin{center}
		\includegraphics[width=0.95\textwidth]{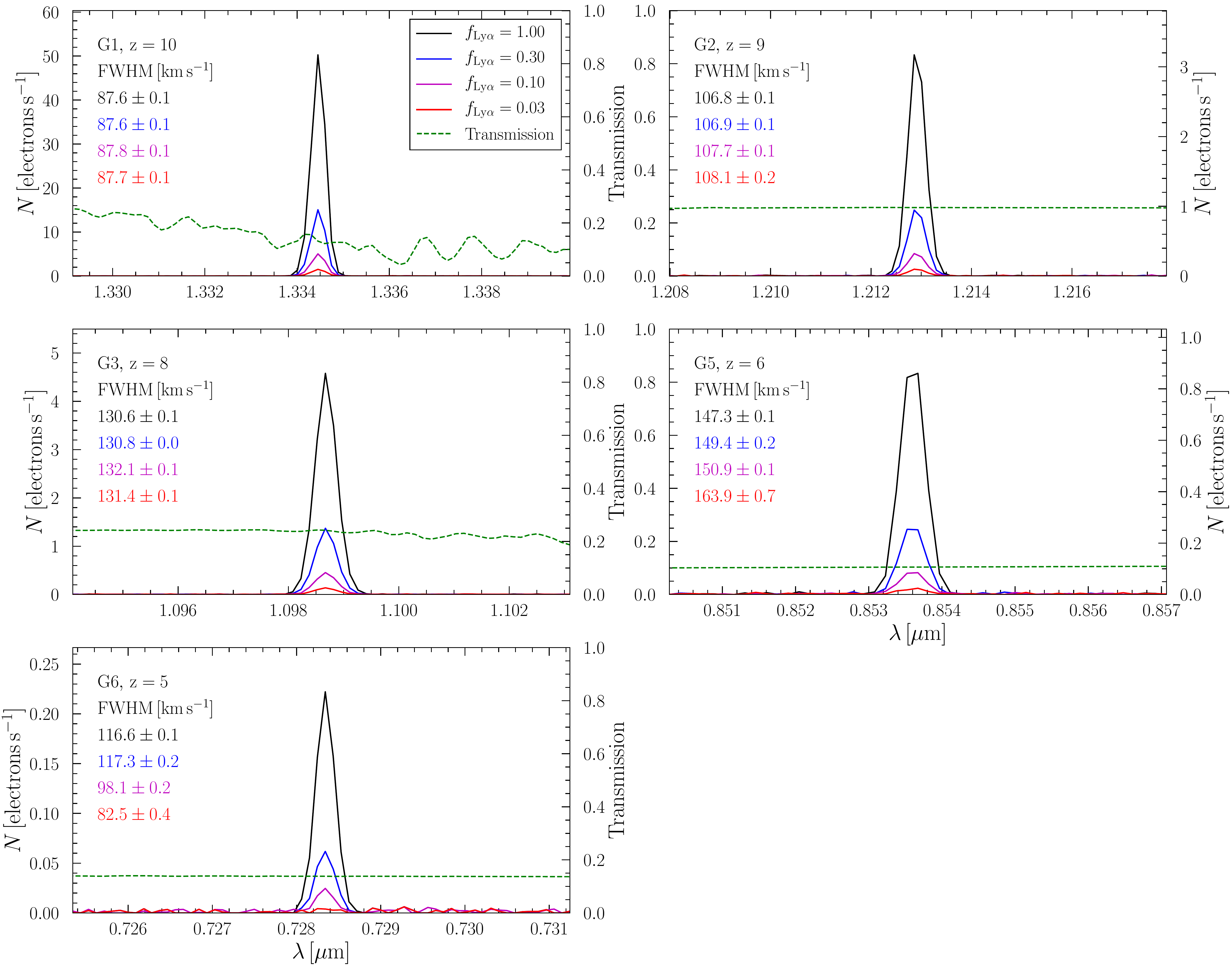}
		\caption{ Single aperture \lya spectra for the eight galaxies after observation wtih \hsim when the ${\rm PopIII.1}$ IMF. For each galaxy we show the unaltered emission line (solid black line) as well as the emission line multiplied by $\flya$ with values of $0.3,\,0.1$ and $0.03$ (solid blue, magenta, red lines respectively). The dashed-black line indicates the wavelength of the peak emission line prior to observation. The green dashed line shows the fraction of light at a given wavelength reaching the detector. Finally each panel states the measured Full Width Half Maximum (FWHM) of each value of $\flya$ (the colour of the text corresponding each line).
		}
		\label{fig:lyalinem1}
	\end{center}
\end{figure*}
\begin{figure*}
	\begin{center}
		\includegraphics[width=0.95\textwidth]{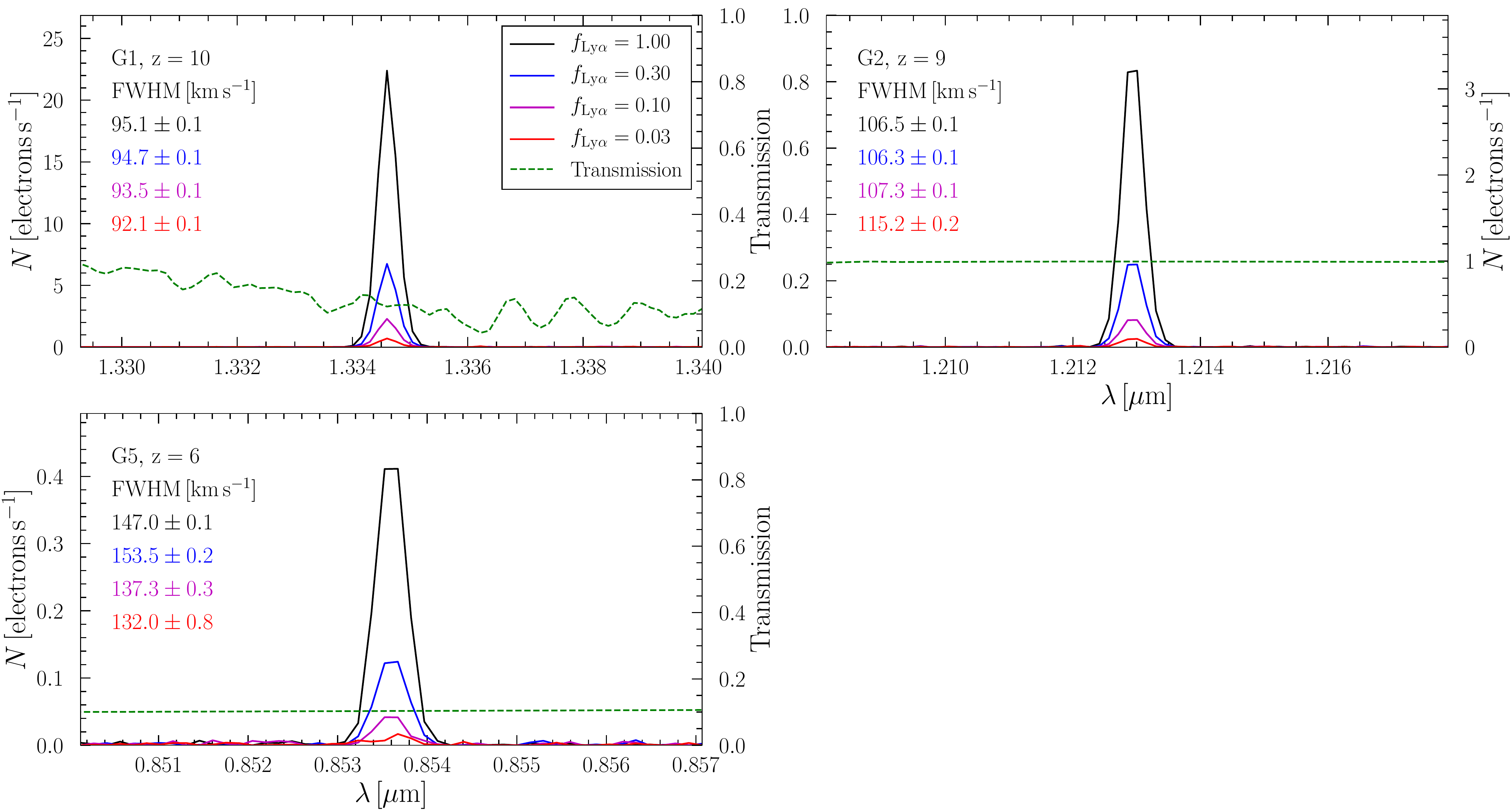}
		\caption{ The same as Fig.~\ref{fig:lyalinem1}, but when the ${\rm PopIII.2}$ IMF is used.
		}
		\label{fig:lyalinem2}
	\end{center}
\end{figure*}

\section{The SED of AGN}
\label{app:agn}

\begin{figure}
	\begin{center}
		\includegraphics[width=0.45\textwidth]{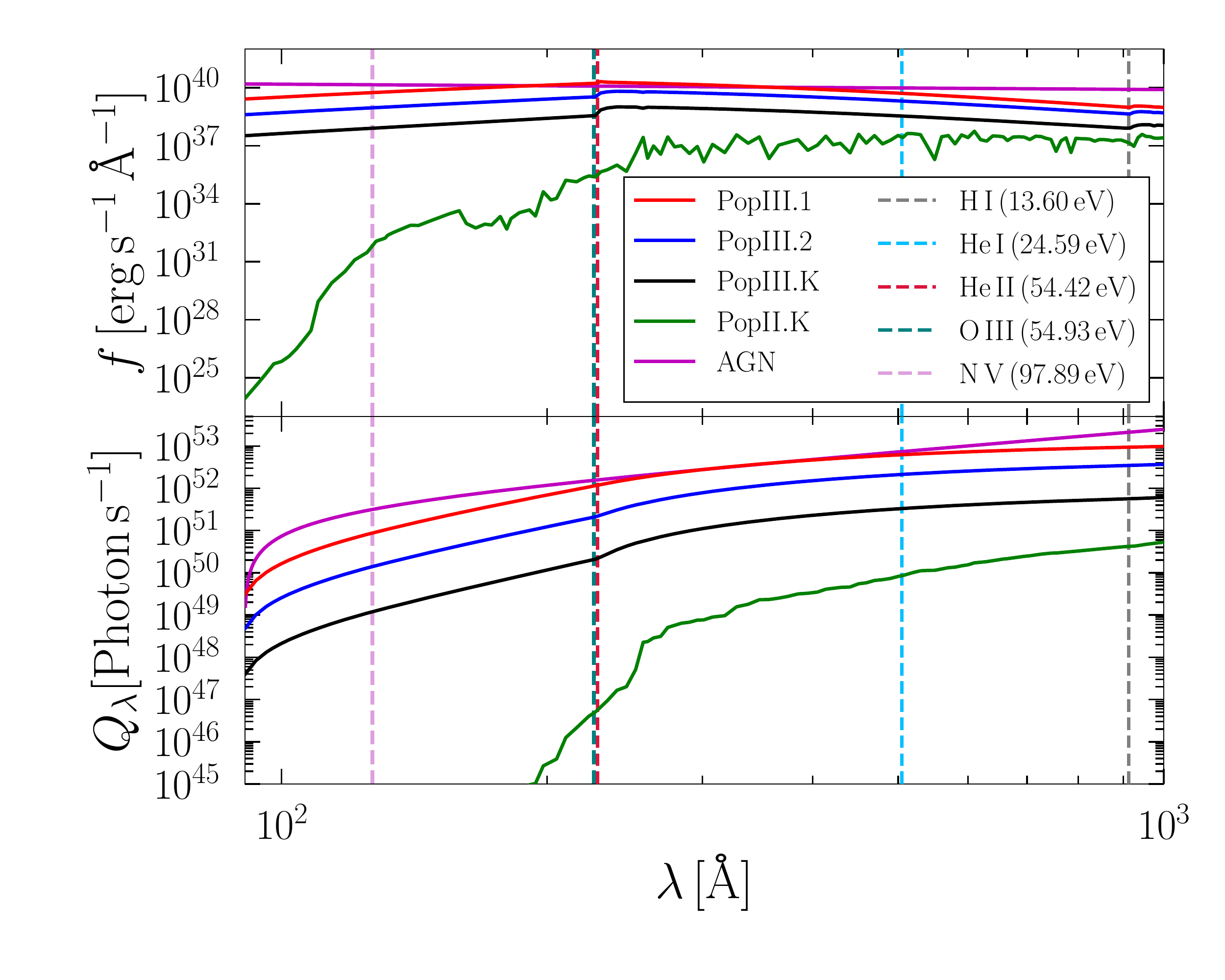}
		\caption{ Comparison of the SED's for PopIII.1, PopIII.2, PopIII.K and PopII.K with an assumed AGN SED. The vertical dashed line show the ionisation potential for H, He, He II, O III and N V  \citep[values are taken from Table 3.5 of ][]{AllensAstro}. 
		}
		\label{fig:agn}
	\end{center}
\end{figure}

The SED of an AGN can be approximated as broken power law given by
\begin{equation}
	f(\lambda) \propto \left\{
	\begin{array}{l l} 
		\lambda^{-0.278} &\quad\mbox{if $\lambda\lesssim 1200\ang$,}\\
 		\lambda^{-1.339} &\quad\mbox{if $\lambda> 1200\ang$,}\\
	\end{array}\right.
	\label{eq:agnsed}
\end{equation}
with a typically luminosity of $\sim10^{12}L_{\odot}$ \citep[see chapter 14 of][]{Osterbrock:2006aa}. The top panel of Fig.~\ref{fig:agn} compares the spectrum given by Eq.~\ref{eq:agnsed} with the SEDs used in this work for \piii stars with an age of $10^{4}\yr$. For each SED we calculate the number of photons that would be able to ionise an atom ($Q_{\lambda}$) with an ionising potential at $\lambda$ (bottom panel of Fig.~\ref{fig:agn}). Therefore by knowing the ionisation potential of an atom, Fig.~\ref{fig:agn} can be used to determine if an AGN or \piii stars should produce a stronger emission line. 

Such an SED produces a large number photons ($\sim 3-5\times10^{53} {\rm s}^{-1}$) with sufficient energies to produce emission lines from the nitrogen V doublet found at $\lambda=1238.82$ and $1242.80\ang$ or the oxygen III emission line at $1665.85\ang$.  \piii stars produce only  $\sim1-30\%$  of the photons that AGN produce at these wavelengths (see bottom panel of Fig.~\ref{fig:agn}) and therefore any resulting nitrogen or oxygen emission lines from a \piii star cluster would be significantly weaker than those produced by AGN. 

It is worth noting that the shape of the AGN SEDs is well constrained at all redshifts \citep{Hao:2014aa}. As a results the values of $Q_{\lambda}$ may vary with $z$ and therefore the impact of AGN on the detection of \piii stars will also vary with redshift.

\begin{table}
	\parbox{0.45\textwidth}{	
		\caption{Number of photons produced by each SED in Fig.~\ref{fig:agn} able to create H II, He II, He III, O IIII [$10^{51}{\rm photon\,s^{-1}}$]}
			\begin{tabular}[h]{l c c c c c}		
			\hline \hline
			Species  & $Q_{\rm AGN}$ & $Q_{\rm PopIII.1}$ & $Q_{\rm PopIII.2}$ & $Q_{\rm PopIII.K}$ &$Q_{\rm  PopII.K}$  \\
			\hline
			H I          &  $211.8$            & $93.4$                     & $34.4$                   & $5.6$                      & $0.42$                    \\
			He I        &  $73.8$              & $61.8$                     & $21.0$                   & $3.3$                      & $0.82$                    \\
			He II       &  $15.8$              & $11.4$                     & $2.1$                     & $0.21$                    & $5.1\times10^{-5}$ \\
			O III        &  $15.5$              & $10.7$                     & $2.0$                     & $0.19$                    & $4.9\times10^{-5}$ \\
			N V         &  $3.2$                & $0.80$                     & $0.13$                   & $0.011$                  & $1.1\times10^{-9}$ \\

			\hline
			\hline
			\end{tabular}\\
		}	

\end{table}

\end{document}